\documentstyle[preprint,revtex]{aps}

\date{September 1992}

\begin{document}
\draft
\preprint{\small UBCTP-92-28, CTP\#2157}
\begin{title}
{\bf Mott Transition in an Anyon Gas}
\end{title}
\author{Wei Chen}
\begin{instit}
Department of Physics, University of British Columbia\\
Vancouver, B.C., Canada V6T 1Z1
\end{instit}
\author{Matthew P.A. Fisher }
\begin{instit}
IBM Research, TJ Watson Research Center \\
PO Box 218, Yorktown Heights, NY 10598
\end{instit}
\author{Yong-Shi Wu}
\begin{instit}
Center for Theoretical Physics, MIT,
Cambridge, MA 02139\\
and Department of Physics, University of Utah\\
Salt Lake City, UT 84112
\end{instit}

\begin{abstract}
We introduce and analyze a lattice model of anyons in a periodic
potential and an external magnetic field which exhibits a
transition from a Mott insulator to a quantum Hall fluid.
The transition is characterized by the anyon statistics, $\alpha$,
which can vary between Fermions, $\alpha=0$,
and Bosons, $\alpha=1$.
For bosons the transition is in the universality class of the
classical three-dimensional XY model.
Near
the Fermion limit, the transition is described by a massless $2+1$
Dirac theory coupled to a Chern-Simons gauge field. Analytic
calculations perturbative in $\alpha$, and also a large N-expansion,
show that
due to gauge fluctuations, the critical properties of the transition
are dependent on the anyon statistics.  Comparison with
previous calcualations at and near the Boson limit,
strongly suggest that our lattice model exhibits a fixed line
of critical points, with universal critical properties which
vary continuosly and monotonically as one passes from
Fermions to Bosons.  Possible relevance to
experiments on the transitions between plateaus in the fractional
quantum Hall effect
and the magnetic field-tuned superconductor-insulator
transition are briefly discussed.

\end{abstract}
\narrowtext

I. INTRODUCTION

A very powerful approach for treating strongly correlated quantum
models in two-dimensions has been to transform the statistics
of the particles by attaching statistical flux tubes \cite{STAT}.
In this way, a fermion model can formally be transformed into
a boson model and vice versa.  Moreover, models describing
particles with fractional statistics (anyons) can be mapped either
way, into a fermion or a boson model. This mapping is usually
accompanied by a flux smearing `mean field' treatment,
originally introduced by Fetter, Laughlin and Hanna \cite{FLH}.
In this mean field approach the statistical flux tubes
are `detached' from the particles and smeared uniformly
in space.  Taken together with the above mapping, this mean
field approach effectively `trades in' statistics for
an external magnetic field! This approach serves as
the key ingredient in various recent theories of the
quantum Hall effect:  For example the so-called
Ginzburg-Landau approaches \cite{GirMac,ZhHK,Read} which focus on
an underlying Bose condensation, the hierarchical construction due
to Jain \cite{Jain,WB} which relates integer and fractional
Hall states, and most recently to theories of the half-filled Landau
level \cite{HLR} as a pseudo-fermi liquid. It has also recently
been applied in attempt \cite{LKZh} to relate the continuous
transitions between plateaus in the fractional Hall effect to
the transition between plateaus in the integer effect.

It is generally believed that this flux-smearing mean field
approximation should be legitimate when the mean field state
is an incompressible fluid with a gap, such as a quantum Hall
fluid.  Indeed, various numerical exact diagonalizations \cite{NumDiag}
on small systems appear to confirm this belief.  However, when
the reference state is gapless, as for example at a transition
between plateaus in the quantum Hall effect or in the
fermi-liquid theories of the half-filled Landau level,
the approximation is much more highly suspect.
In particular various conclusions arrived at in Ref. \cite{LKZh}
which relate detailed critical properties at transitions
between different quantum Hall plateaus are worth detailed scrutiny.

The purpose of the present paper is to introduce the simplest
possible anyon model which exhibits a continuous phase
transition, and analyze in detail the universal critical
properties near and at the transition. The model is
characterized by the statistics of the particles,
which can be tuned continuously from Bose to Fermi.
The particles are subject to a periodic potential with a period
commensurate with the particle's density, corresponding
to one particle per unit cell.  Moreover, the model
includes an external magnetic field, taken as zero in the
boson limit, one flux quantum per particle in the fermion limit,
and proportional to the statistics in the anyon case. In this way
a `mean-field' treatment of the model, as described above,
would result in behavior fully independent of the
particle's statistics. Controlled calculations of the actual
critical properties then allow for a direct check on the
validity of the mean-field approach.

In the boson limit, the model exhibits a Mott-insulator to
superfluid transition which is in the universality class
of the classical 3D XY model \cite{FWGF}.
Critical properties can be
extracted via Monte Carlo simulations or 1/N expansions
\cite{MCXY}.
As we show below, in the Fermion
limit the model is soluble and exhibits a `gap-closing'
transition between a band-insulator and an integer quantum Hall
state.  We analyze the critical properties in this case,
which are described by a massless 2+1 Dirac equation,
and find that they are most certainly
different from the 3D XY model, which reveals the
inadequacies of the `mean-field' approximation.

For anyon statistics the model exhibits a transition between
a Mott-insulator and a quantum Hall fluid.  Can the critical
properties be extracted for this anyonic Mott transition?
As shown recently by Wen and Wu \cite{WW},
it is possible to perform a renormalization group (RG) calculation
perturbative in deviations of the particle's statistics
from the bosonic point.
Within a controlled 1/N expansion they
find that the critical properties vary continuously with the
particle statistics! In RG terminology they have found
a `fixed line' parameterized by the particle's statistics.
In this paper we perform a similar perturbative analysis
expanding around the Fermi end, described by the Dirac equation.
We likewise find that the critical exponents and other universal
properties vary continuously with statistics. Again, this
statistics-dependence is shown to originate from gauge-field
fluctuations near the critical point.  Moreover the sign of our
perturbative results for the critical exponents
near the Fermion point, suggest that
the exponents vary monotonically upon moving along the
fixed line between the Fermion and Boson transitions.
We suspect that this might be a generic feature of
quantum phase transitions involving fractional statistics
particles.  This would imply that the
leading perturbative
corrections to the flux-smearing mean field treatment approach
monotonically the exact behavior.

The dependence of the detailed critical properties
on the statistics of the underlying particles in our model system,
strongly suggests that, in contrast to recent speculations \cite{LKZh},
the phase transition between integer plateaus in the
quantum Hall effect is not in the same universality class
as that between fractional plateaus.  Indeed, in view of our results,
it would be much more natural to argue that the exponents will
depend on the statistics of the condensing quasiparticles,
which of course depend on which two fractions the transition is between.
However since both disorder and long-ranged Coulomb interactions
are relevant perturbations at the Mott transition studied here,
we cannot make any definitive statements about the universality class
appropriate to experimental quantum Hall phase transitions.

The paper is organized as follows.  In Section II we
introduce the lattice anyon model, and discuss
the continuum limit, with special emphasis on the fermion limit.
The resulting critical theory is a massless 2+1 Dirac equation
coupled to a Chern-Simons term. Section III is
devoted to a detailed RG analysis of this fermionic Chern-Simons
theory, perturbative in deviations of the statistics away from
Fermi, as well as in a large N limit.  In addition to
the critical exponents $\nu$ and $\eta$ we calculate the universal
Hall and longitudinal conductivities.  Finally, Section IV is
a discussion section with emphasis on possible relevance
to experimental systems.

II. THE MODEL

In this Section we introduce our lattice anyon model,
and in the fermion case extract the appropriate continuum
limit.

The Hamiltonian we study is defined for convenience in terms of
spinless fermion operators:
\begin{equation}
H= - \sum_{ij} ( t_{ij} c_i^\dagger c_j + h.c. )
+ m \sum_j s_j c_j^\dagger c_j\;.
\label{1}
\end{equation}
where the fermion operators satisfy the usual anticommutation
relations:
\begin{equation}
\{c_i , c_j^\dagger \} = \delta_{ij}\;. \nonumber
\end{equation}
Here i and j label sites of a square lattice, $t_{ij}$ is
a hopping matrix element and $s_j=+1$ for sites of one sub-lattice
and $-1$ on the other sub-lattice.  The second term in the Hamiltonian
is thus a staggered potential with strength $m$.
We restrict attention exclusively to half-filling, so that the
number of particles equals the number of sites on either of the two
sub-lattices.

An applied magnetic field, B, and flux tubes attached to the
fermions both enter via a gauge field: $t_{ij} = | t_{ij} |
e^{iA_{ij}}$. For later convenience we take the hopping
strength $|t_{ij}|$ to be equal to $t$ for nearest neighbor
sites, $t'/4$ for next nearest neighbors and zero otherwise.
The gauge field is chosen so that
\begin{equation}
\nabla \times A = B + \alpha \rho\;,
\label{2}
\end{equation}
where $\nabla \times A$ denotes a lattice curl, that is an oriented
sum of $A_{ij}$ around plaquettes, and B is a uniform external
field in units of a flux quantum, $h/e$, per elementary square
of the lattice.  Here $\rho$, which is an operator,
denotes the particle density so that Eq.(3) must be thus taken
as a constraint. The statistics parameter
$\alpha$ gives the strength of the statistical flux tubes
which are thereby attached to each particle.  With $\alpha =1$
the particles are transmuted into bosons, whereas anyons
correspond to $\alpha$ between zero and 1.
Finally, to fully specify the model we choose the external field
as $B=(1-\alpha)/2$, so that at half-filling
in a flux-smearing mean-field treatment the
Hamiltonian reduces to bosons in zero field, for all $\alpha$.

Notice that the model has been constructed so that
in the boson limit ($\alpha =1$) it is formally equivalent
to a lattice model of hard core bosons in zero magnetic field.
Moreover, at half-filling the boson density is
commensurate with the period of the staggered potential.
This boson model is expected to undergo
a Mott-insulator to superfluid transition \cite{FWGF}
as the ratio of the kinetic energy ($t$, $t'$) to
the staggered potential ($m$) is varied.
This zero temperature quantum transition is expected to
be in the universality class of the classical 3-dimensional XY model.
The appropriate coarse-grained continuum theory to describe
the behavior near the transition is then simply
a $\phi^4 $ complex scalar field theory.

For $\alpha$ near 1, the model corresponds to anyons with
statistics `close' to bosonic, in a weak magnetic field.
In this case the model is expected to exhibit a transition
from a Mott-insulator to a gaped quantum Hall fluid.
The transition can be studied by adding a Chern-Simons term
to the scalar field theory, which attaches $(1-\alpha)$
flux tubes to each boson.  The corresponding critical behavior
has been studied in a recent large-$N$ approach \cite{WW},
and the exponents found to vary continuously with $\alpha$.

When $\alpha=0$ the Hamiltonian is simply that of non-interacting
spinless fermions in a magnetic field, and can be solved by
straightforward diagonalization.  Specifically,
with 1/2 flux per square and the staggered potential the unit cell
has four sites \cite{fluxphase},
as sketched in Fig.~$1$.  Extracting the band
structure thus involves diagonalizing a 4 by 4 matrix.
Doing so reveals that generally there are two bands,
symmetric about zero energy (half-filling), with a band gap
about the origin.  Depending on the relative sizes of
$t'$ and $m$ the bands can be either `Landau levels', which
contribute a Hall conductance, or conventional insulating band
with no Hall effect \cite{TKNN}.
Of interest to us is the transition between
these two phases, where the band gap closes. (Across the transition
the Hall conductance jumps from unity to zero, in units of
$e^2/h$ with $e$ the electric charge of the fermion.)
The band structure reveals that the gap closes
at a single point in crystal momentum ($k$-) space.
By focusing on values of $k$ near this point,
we now construct a continuum field theory for the behavior
near the transition.

To this end, we choose an explicit gauge for the
$A_{ij}$, which is shown in Figure (1).  Moreover,
we put $t$=1 and take $t'$ and $m$ much smaller than 1.
In this limit the band gap vanishes when $t' = m$
and occurs at $(\pi, \pi)$ in the Brouillon zone.
Upon linearizing the momentum about this point
the Hamiltonian can be cast in the form:
\begin{equation}
H = \int_{p} \sum_{a.b=1}^{4} c_a^\dagger (p) H_{ab} (p) c_b (p)\;,
\label{3}
\end{equation}
where $c_a (p)$ denotes the Fourier transform of
the electron operators on the `$a$' sites (see Fig.~$1$) at
momentum $(\pi , \pi ) + p$.
The 4 by 4 matrix $H_{ab}$ can be written \cite{SHANKAR}:
\begin{equation}
H_{ab} (p) = (p_y \sigma_y - m \sigma_z ) \otimes
1 + (p_x \sigma_x + t' \sigma_z )
\otimes \tau_y\;,
\label{4}
\end{equation}
where $\sigma_\mu$ and $\tau_\mu$ denote the usual Pauli matrices and
$\otimes$ is a direct product.  Upon performing a basis rotation on
$\tau_\mu$ by 90 degrees about the x-axis, the Hamiltonian
can be transformed into a 2 by 2 block diagonal form,
\begin{equation}
H_{ab} (p) = H_+ (1+\tau_z )/2  +  H_- (1-\tau_z)/2\;,
\label{5}
\end{equation}
with
\begin{equation}
H_{\pm} = \pm p_x \sigma_x + p_y \sigma_y + (\pm t' -m ) \sigma_z\;,
\label{6}
\end{equation}

Notice that $H_+$ is a two-dimensional Dirac Hamiltonian
with mass, $M=t'-m$, which vanishes at the transition.
Under $\sigma_x \rightarrow -\sigma_x$, $H_-$ is also
a Dirac Hamiltonian except with mass $t'+m$.  This mass
remains non-zero when $m=t'$, and thus non-critical.
As far as critical properties near the transition are
concerned we can safely ignore the massive field and
focus exclusively on the single Dirac field which
goes massless.  Since this theory is non-interacting
we can easily extract all of the relevant critical
properties (see Section 3).

When the statistics parameter $\alpha$ is non-zero, but small,
progress can be made by expanding around the fermion point
(ie. the Dirac equation).  The appropriate continuum theory
near the transition in this case can be simply obtained
from the Dirac equation by minimal coupling to the gauge
field whose associated flux must be attached to the
fermions.  This can be achieved in the usual way by the
addition of a Chern-Simons term \cite{CSL} to the 2+1
Dirac Lagrangian. In the next Section we
perform a renormalization group analysis
on this 2+1 Dirac plus Chern-Simons theory to extract
critical properties near the anyon Mott transition.
Two expansions are available for calculating them: the weak
`coupling' expansion, in which the small parameter is $\alpha$,
the deviation from Fermi statistics, and the $1/N$ expansion,
if one consider the case with $N$ species of anyons.

III. CRITICAL PROPERTIES

The quantities of primary physical interest which characterize
the Mott transition are the transverse and longitudinal
conductivities
at the critical point, and the usual
critical exponents $\eta$ and $\nu$.
It has recently been argued \cite{FGG}
that in general at zero temperature
quantum phase transitions in two spatial dimensions,
the conductivity should be universal, and we indeed
verify this below for the anyon Mott transition.
As usual we define the exponent $\nu$ in terms
of the correlation length which diverges as
$M^{-\nu}$, upon approaching the transition by taking
the Dirac mass $M=m-t^\prime$ to zero.
We define the exponent
$\eta$ via the decay
of the two-point anyonic correlation function
\begin{equation}
<\psi^\dagger (x) \psi(0)> \sim x^{-(1+\eta)} \; ,
\label{etadef}
\end{equation}
where $\psi$ is the anyon field operator in either
a fermionized or bosonized representation.
Implicit in this definition is the assumption that
the anyons are converted into Fermions or Bosons
by attaching flux tubes in the Landau gauge, where the
vector potential is divergence free.
Notice that
in the above definition of $\eta$ we have pulled out a factor of 1,
so that it coincides
with the usual exponent $\eta$ at the Boson Mott-insulator
to superfluid transition \cite{FWGF},
where the 1 is twice the canonical dimension
of the boson field and $\eta$ twice the anomalous dimension.
However, as we see below, at the Fermion Mott transition
the scaling dimension of the Fermi field is 1, and there is no
anomalous dimension, so that in this case $\eta=1$, rather than
twice the anomalous dimension.

To extract critical properties near the Fermion point we employ
a renormalization group (RG) analysis
in the framework of continuum Euclidean field theory \cite{Amit,Parisi}
by applying standard
diagrammatic perturbative techniques \cite{Ramond}.

1) Field Theory Approach

As shown in Section II, the long-length scale and low energy physics
near the anyon Mott transition can be described by two
D=2+1 Dirac fermions, one of which becomes massless at the
transition, which are coupled to a
Chern-Simons gauge field $a_{\mu}$.  Specifically, from the
Hamiltonian $H$ in Eq. (\ref{5}) and the constraint
in Eq. (\ref{2}),
the appropriate
Euclidean Lagrangian is simply
\begin{equation}
L= \sum_\pm \psi_{\pm}^{\dagger} [\gamma_\mu (\partial_\mu - ig a_\mu)
+iM_\pm ]\psi_\pm
+i{1\over 2} \epsilon_{\mu\nu\lambda}
a_\mu \partial_\nu a_\lambda\;.
\label{action1}
\end{equation}
where $\psi_\pm$ are the two species of Dirac fields.
Here the Dirac matrices in three dimensions
are $\gamma_\mu=
i\sigma_\mu$, with $\sigma_\mu$ $(\mu=1,2,3)$ Pauli matrices,
so that
\begin{eqnarray}
\gamma^\mu \, \gamma^\nu = - \delta^{\mu\nu}{\bf 1}
 - \epsilon^{\mu\nu\lambda}\gamma^\lambda\;, \;\;\;\;\;
{\bf Tr}(\gamma^\mu) = 0\;.
\end{eqnarray}
We have normalized the coefficient of the Chern-Simons term,
and the `coupling constant' $g^2$ is essentially the anyon
statistics, $\alpha$, measured from the fermion point:
\begin{equation}
g^2 = 2 \pi \alpha\;.
\label{coeff}
\end{equation}
A source electromagnetic field which can be used to
calculate the conductivites within linear response
is minimally coupled to the Dirac fields in the usual way.
It is worth emphasizing that the sign of the Chern-Simons
term in the Lagrangian (9) must be chosen correctly in order
that the flux tubes attached to the particles are
in the opposite direction to the external physical magnetic field.

The phase transition between the quantum Hall phase and the Mott
insulator occurs when one of the Dirac masses, say $M=M_+$,
passes through zero.  Before discussing the associated critical
properties it is instructive to first recover the expected behavior
of the two phases from the above Lagrangian.
Provided we put in an ultraviolet cutoff, which is appropriate
for the original lattice theory, when both masses are non-zero
straightforward perturbation theory in powers of $g^2$ is
convergent.  Consider, for example, the fermion
`polarization tensor', defined as the Fermion
current-current correlation function,
which we denote as $\Pi_{\mu\nu}(p)$. Current
conservation (or gauge invariance) implies that the polarization
tensor is always transverse: $p_\mu \Pi_{\mu\nu}(p)=0$. Such a tensor
in $D=3$ can be decomposed into an even and odd part:
\begin{equation}
\Pi_{\mu\nu}(p)=
\Pi_e(p) p
(\delta_{\mu\nu}-{p_\mu p_\nu\over p^2})
+ \Pi_o(p) \epsilon_{\mu\nu\lambda}p_\lambda\;.
\label{pi}
\end{equation}
where we have used the notation $p=|p|$. From standard
linear response theory, the two parts
are related to the Fermion longitudinal
and transverse conductances, in units
of $e^2/h$ with $e$ the charge, by
\begin{equation}
\sigma_{F,xx} = -{1 \over \alpha} \Pi_e(0)\;, \;\;\;
\sigma_{F,xy} = -{1 \over \alpha} \Pi_o(0)\;.
\label{condu1}
\end{equation}
The conductance of the anyons has an additional contribution
coming from the attached flux tubes.
Formally this is due to the difference between the anyon current
operator,
$\gamma_\mu \psi^\dagger (\partial_\mu - iga_\mu ) \psi$,
which couples to the physical electromagnetic field,
and the Fermion current operator:
$\gamma_\mu \psi^\dagger \partial_\mu \psi$.
(see Fig.~$2$).
As can be easily shown (see e.g. \cite{WW}),
the anyon resistivty tensor, $\rho_{ij}$,
is simply related to the fermionic resistivity tensor via:
\begin{eqnarray}
\rho_{ij} = \rho_{F,ij} - \epsilon_{ij} \alpha .
\label{condu2}
\end{eqnarray}

Previously there has been an intense effort in analyzing
the
Chern-Simons gauge theory coupled to
a Dirac field
in perturbation theory employing standard
diagrammatic techniques
\cite{DJT,CH,NSW,SSW,ChSW1,ChSW2,Ch1,Ch2,Other}.
Specifically, for a massive Dirac field it has been shown
that the one-loop diagram
in Fig.~$3a$ contributes
a finite contribution to the odd part of the
Fermion polarization, $\Pi_o^{(1)} (p=0) = + (\alpha/2) sgn(M_\pm)$.
Notice the dependence on the sign of the Dirac mass.
Moreover it has been proved \cite{CH,SSW}
that in the massive theory there are
no additional non-zero corrections to the Fermion polarization,
$\Pi_{e,o} (0)$,
to all orders in $\alpha$.
Thus away from the Mott transition the Fermion conductivity
tensor which follows from Eqn. \ref{condu1} is given by
\begin{equation}
\sigma_{F,ij} = - {1 \over 2} \epsilon_{ij} \sum_\pm sgn(M_\pm)  .
\label{condmass}
\end{equation}
In the Mott insulating phase, where the two masses have different
signs, the conductivity tensor vanishes as expected,
whereas in the Quantum Hall effect phase one has
simply $\sigma_{F,ij} = - \epsilon_{ij} $.
The total anyon resistivity tensor in the quantum Hall phase
is then $\rho_{ij} = \epsilon_{ij} (1-\alpha)$.
Notice that when $\alpha=1$, which corresponds to the bosons
superfluid phase (in zero magnetic field),
the Hall resistivity vanishes as expected.

We now turn to the critical properties near the transition
where the mass of one Dirac field, $M=M_+$ vanishes.
In the following we focus exclusively on the massless Dirac
field, since the contributions from the massive field
are known to all orders, as described above.
Consider a renormalization group (RG) transformation
for this massless field, which involves integrating over a
shell of (three-)momenta $p$ in a shell between $\Lambda$
and $\Lambda/b$, with $b>1$.  To complete the RG we
rescale the momenta by $b$ and
the fields $\psi$ and $a$ by
$b^{1+\gamma_\psi}$ and $b^{1+\gamma_a}$, respectively.
The anomalous dimensions will be chosen to keep
the coefficients of the quadratic terms in the Lagrangian Eqn.
(\ref{action1})
fixed.  Power counting at the Gaussian ($g=0$)
fixed point reveals that the coupling constant $g$ is indeed
dimensionless.  A straightforward perturbative renormalization
group can then be carried out using standard methods.
As defined above in Equation (\ref{etadef})
the critical exponents
$\eta$
and $\nu$ are related to the anomalous
dimension of the massless Fermion field, $\psi$, and that of the
composite operator $\bar{\psi}\psi$, which we denote by
$\gamma_{\bar{\psi}\psi}$
\cite{Amit,Parisi}:
\begin{equation}
\eta = 1 + 2 \gamma_{\psi}\;, \;\;\;\;
\nu^{-1} = 1 - \gamma_{\bar{\psi}\psi}
\label{expid}
\end{equation}

It can be easily shown (\cite{SSW}) at one-loop order that the
beta-function, $\beta(g) = -dg/dln(b) = O(g^5)$, vanishes.
Indeed it has been suggested \cite{SSW} and shown explicitly to second
order, that the beta-function vanishes identically:
\begin{equation}
\beta(g)=0\; .
\end{equation}
This implies that there is a line of
fixed points or critical points, parametrized
by the anyon statistics $\alpha$!
Unfortunately all of the anomalous dimensions vanish at one-loop
order, so it is necessary to go to second order
to evaluate the leading non-trivial corrections to the
exponents.
In addition to the critical exponents, the universal
anyon conductivity at the critical point is of interest.
As first emphasized by
Semenoff, Sodano and Wu \cite{SSW}, in contrast to the massive theory,
in the massless theory there is a non-vanishing two loop contribution
to the conductivity.  Indeed,
the exact value of this
correction was obtained subsequently
by Chen \cite{Ch1,Ch2}.
In the following we will give a unified description of the
two-loop perturbative results for the critical properties
of the model (\ref{action1}) in the context of the anyon
Mott transition, together with some new calculations when
the quantities of interest are not available in the literature.

2) Weak-Coupling Expansion

In this subsection we carry out a two-loop
perturbation expansion in powers of
the coupling constant $g^2$ to extract critical exponents and
the universal conductivity.
We take the Landau gauge, by
adding a gauge fixing term,
$(1/2\lambda) (\partial_\mu a_\mu)^2$, to the Lagrangian
(\ref{action1}) and then take the limit $\lambda
\rightarrow 0$ in the resulting propagator for the
gauge field.  With the normalization chosen as shown in
eq.(\ref{action1})
the appropriate Feynman rules for the massless Dirac field
in the Landau gauge are
\begin{eqnarray}
& & {\rm the~ fermion~ propagator}\;\;\;\;\;
S_0(p) = {1\over i\not\!p}\;,\\
& &{\rm the~~ gauge~ propagator}\;\;\;\;\;
G_0^{\mu\nu}(p) = -{\epsilon^{\mu\nu\lambda}p^\lambda\over p^2}\;,
\\
& &{\rm the~~ interaction~ vertex}\;\;\;\;\;
\Gamma_0^\mu (p) = ig\gamma^\mu\;.
\end{eqnarray}
The advantage of the Landau gauge is that no infra-red divergences
would appear perturbatively in this gauge \cite{DJT,SSW,ChSW1,ChSW2}.

Since we require a two-loop calculation,
rather than regularizing in the ultraviolet with a
finite momentum cutoff,
$\Lambda \approx 1/a$, it is more
convenient to
regularize by dimensional reduction and also employ a minimal subtraction
method.
In this approach, which was first
suggested and extensively used in Refs. \cite{ChSW1,ChSW2}
for the $D=3$ Chern-Simons gauge theory,
all momentum integrals in the loop-expansion are continued
to general dimension D:
\begin{equation}
\int {d^3k \over (2\pi)^3} \rightarrow
\mu^{3-D} \int {d^D k \over (2\pi)^D}\; ,
\label{dimen}
\end{equation}
where, in order to balance the dimension, one must introduce a
parameter $\mu$ which has dimension of mass.
All vector, tensor and spinor quantities and
in particular the symbol $\epsilon^{\mu\nu\lambda}$ that appears
in the original Feynman integrands, though,
are always treated as if
they were formally three dimensional. This implies that we will use
the identities
\begin{eqnarray}
\delta^{\mu\mu} &=& 3\; , \\
\epsilon^{\mu\nu\lambda}\epsilon^{\mu\tau\eta} &=&
\delta^{\nu\tau}\delta^{\lambda\eta}-\delta^{\nu\eta}
\delta^{\lambda\tau}\;
\end{eqnarray}
before performing any momentum integrals.  Finally,
we adopt a minimal subtraction method for renormalization,
by removing the simple poles in
$\epsilon\equiv 3-D$ \cite{NoteB} and setting all higher-order
poles to be zero.
The parameter $\mu$ we introduced in
Eqn. (\ref{dimen}) represents, as usual, the ``renormalization point''
in the minimal subtraction approach.
In all cases that have been
checked \cite{ChSW1,ChSW2,Ch1,Ch2}, including
pure Chern-Simons
(Abelian or non-Abelian) gauge theory, coupled or not to
massless or massive boson or Fermion
matter fields, this approach to
regularization and renormalization
has been shown to respect both
gauge invariance and Ward identities,
at least up to two loops.

We consider first the universal conductances
at the critical point ($M=0$), focussing exclusively on the
contribution from the massless Dirac field.
(The contribution from the massive field is given in
Eqn. (\ref{condmass}).)
The one-loop polarization
tensor $\Pi_{\mu\nu}^{(0)}(p)$, due to the (massless)
fermion bubble
(Fig.~$3a$), is purely symmetric in the indices $\mu$
and $\nu$ and given by:
\begin{equation}
\Pi_e^{(1)}(p) = - {g^2\over 16} \;, \;\;\;\;\;
\Pi_o^{(1)}(p) = 0\;.
\label{pi1}
\end{equation}
Thus upon using Eqn. (\ref{coeff}) and (\ref{condu1}), to
leading (zero'th) order in $\alpha$,
the total Fermion (and anyon) conductivity at the Mott transition
is given by
$\sigma_{F,xy} = 1/2$, coming solely from the massive Dirac field,
and $\sigma_{F,xx} = \pi /8$.
It is interesting that this value is precisely equal to the
value of $\sigma_{xx}$ obtained at the Boson Mott-insulator
to superfluid transition within a large-N limit \cite{FGG}.

The leading (linear)
corrections in $\alpha$ come from the three
two-loop diagrams, shown in Fig.~$4$, which
contribute to the polarization tensor.
It turns out that all three contributions are
antisymmetric in the indices $\mu$ and $\nu$, namely:
\begin{equation}
\Pi_{\mu\nu}^{(2)}(p)=\epsilon_{\mu\nu\lambda}p_\lambda\;
\Pi_o^{(2)}(p)\;.
\end{equation}
This implies that there is no correction at this order to the
longitudinal Fermion conductivty, $\sigma_{F,xx}$.
The contribution to the Hall conductivty can be obtained
by contracting each diagram in Fig.~$4$ with
${1\over 2p^2}\epsilon^{\mu\nu\lambda}p^\lambda$
and performing a trace over the Dirac spinor-components, which gives:
\begin{eqnarray}
\Pi_{o(a+b)}^{(2)} &=& 8{g^4\over p^2}\mu^{6-D^\prime-D}\int
{d^{D^\prime}k\over (2\pi)^{D^\prime}} {d^Dq\over (2\pi)^D}
{p\cdot(k+p)q\cdot(q+k)
\over k^2(k+p)^2(k+q)^2q^2}\;,
\label{abx}\\
\Pi_{o(c)}^{(2)} &=& 2{g^4\over p^2}\mu^{6-D^\prime-D}\int
{d^{D^\prime}k\over (2\pi)^{D^\prime}} {d^Dq\over (2\pi)^D}
{F(q,k,p)
\over k^2(k+p)^2(k+q)^2(k+q+p)^2q^2}\;,
\label{cx} \\
F(q,k,p) &=& q^2{\bf [}3k^2p^2-2(k\cdot p)^2+(k\cdot p)p^2
             -(k\cdot p)(q\cdot p) + (k\cdot q) p^2{\bf ]}\nonumber\\
& & -k^2(q\cdot p)^2
   + (k\cdot q){\bf [}2(k\cdot p)(q\cdot p) -2(k\cdot q)p^2
   -(q\cdot p)p^2{\bf ]}\;.
\end{eqnarray}
Since the integrals over
$k$ in Eqs.\ (\ref{abx}, \ref{cx}) are convergent directly in
$D^\prime = 3$ dimensions, we perform them there to
obtain
\begin{eqnarray}
\Pi_{o(a+b)}^{(2)} &=& {g^4\over 4p^2}\mu^{(3-D)}\int
{d^Dq\over (2\pi)^D}
{\bf [}{p\over q|q+p|} - {p\cdot(q+p) \over q^2|q+p|}{\bf ]}\;,
\label{abxx}\\
\Pi_{o(c)}^{(2)} &=& {g^4\over 4p^2}\mu^{(3-D)}\int
 {d^Dq\over (2\pi)^D}
{\bf [} {q\cdot p|p|\over q^2(q+p)^2}
+{p\cdot(q+p)\over q^2|q+p|}{\bf ]}\;.
\label{cxx}
\end{eqnarray}
The second terms in Eqs.\ (\ref{abxx}, \ref{cxx}) are each
logarithmically
divergent in $D=3$ but cancel one another exactly.
The remaining finite contribution in $D=3$ is finally:
\begin{equation}
\Pi_o^{(2)}(p)=\Pi_o^{(2)}(0)=\Pi_{o(a+b)}^{(2)}
+\Pi_{o(c)}^{(2)} = -{g^4\over 16\pi^{2}}(1+{\pi^2\over 4})\;.
\label{piox}
\end{equation}
This result was obtained previously in Ref.
\cite{Ch2} and, up to an overall sign, independently in
Ref. \cite{ST} (where the convention is that there is no imaginary unit
in front of the Chern-Simons term).

Upon inserting Eqn.(\ref{pi1}) and (\ref{piox}) into
(\ref{condu1}) we obtain
the final result for the Fermion
conductivty at the anyonic Mott transition,
valid up to
first order in $\alpha$:
\begin{eqnarray}
\sigma_{F,xx} &=&  {\pi\over 8 }  \;, \\
\sigma_{F,xy} &=& - {1 \over 2} +
{1\over 4}(1+{\pi^2\over 4})\alpha  \; .
\end{eqnarray}
The contribution of -1/2 at $\alpha=0$ comes from the
second Dirac field which remains massive at the transition.
The anyon conductivity follows by inverting to get the Fermion
resistivity, and then using Eq. (14).
Note again that at this order there is no
correction to the longitudinal Fermion conductivity.
An advantage of the 1/N method, which we describe in the
next section, is that non-trivial corrections to
$\sigma_{F,xx}$ do appear at leading order in 1/N.

To obtain the critical exponent $\eta$, let us consider the
fermion self-energy $\Sigma(p)$, defined as usual via the
full Fermion propagator:
\begin{equation}
S(p)^{-1} = S_0(p)^{-1} - \Sigma (p) \;.
\end{equation}
 At one-loop order it is given by the diagram in
Fig.~$3b$:
\begin{equation}
\Sigma^{(1)}(p) = -i{g^2\over 8}p\;.
\end{equation}
The two-loop fermion self-energy diagrams are given in Fig.~$5$.
In Ref. \cite{ChSW2} these diagrams have been evaluated to obtain:
\begin{equation}
\Sigma^{(2)}(p) = i\not\!p \{ {g^4\over 24\pi^2}
{\bf [}{2\over 3-D}+ln({\mu^2\over p^2}){\bf ]}+ finite\}\;.
\end{equation}
Notice that at this order
the renormalized mass remains zero, $\Delta M =
\Sigma(p=0)=0$, implying that conformal invariance survives
quantum fluctuations at the transition point $M=0$.
The Fermion wave function renormalization constant is
extracted to be
\begin{equation}
Z_\psi^{-1}= {\bf [} {\partial S^{-1}(p) \over
\partial (i\not\! p)} {\bf ]}_{p=0}
= 1 - {g^4\over 12\pi^2}{1\over 3-D}\;.
\label{zpsi}
\end{equation}
One can thereby obtain
the anomalous dimension $\gamma_\psi$ of the massless
anyon field up to second order:
\begin{equation}
\gamma_\psi
= - {1\over 2}{\partial Z_{\psi}\over \partial (1/\epsilon)}
= - {g^4\over 24\pi^2}\; .
\label{anom1}
\end{equation}

Finally let us consider the renormalization of the
composite operator $\bar{\psi}\psi$, from which we can
extract the critical exponent $\nu$. We insert this
composite operator into one- and two-loop fermion
self-energy diagrams, as shown in Figs.~4 and 5,
with the insertion represented by a cross. To simplify
the calculation,  the external momentum associated with
the $\bar{\psi}\psi$ insertion is taken to be zero.
We start with the one-loop diagrams, i.e. Fig.~$6a$, $6b$, and $6c$.
Within our
regularization scheme these
diagrams are finite in $D=3$:
\begin{eqnarray}
(6a) = (6b) &=&
     -i{g^2\over 8}{\epsilon^{\mu\nu\lambda}p^\lambda \over p}\;,\\
(6c) &=&
       -{g^2\over 8}{\not\!p\over p}\;.
\end{eqnarray}
so that the anomalous dimension of
the operator $\bar{\psi}\psi$ indeed vanishes at one-loop order.

However, the two-loop diagrams in Fig.~$7$ have logarithmic
divergences. After lengthy but straightforward calculations
we have (up to finite contributions which do not contribute to
the anomalous dimension)
\begin{eqnarray}
(7a_1) &=&
       {g^4\over 8}\mu^{3-D}\int {d^Dk\over (2\pi)^D}
            {1\over k(k+p)^2}\; ,\nonumber\\
(7a_2) + (7a_3) &=&
       {g^4\over 2}\mu^{3-D}\int{d^Dk\over (2\pi)^D}
                        {1\over k(k+p)^2}\;, \nonumber\\
(7b_1) &=& -{g^4\over 4}\mu^{3-D}\int
                       {d^Dk\over (2\pi)^D}
                        {1\over k(k+p)^2}\;,\nonumber\\
(7b_2) + (7b_3)  &=& {g^4\over 2}\mu^{3-D}\int{d^Dk\over (2\pi)^D}
                         {1\over k(k+p)^2}\;.
\end{eqnarray}
The calculation of Figs.~$7c_1-7c_3$ is a bit more complicated,
because of the Dirac matrices. However, since we are only
interested in the contribution
proportional to the $2\times 2$ unit matrix
${\bf 1}$, it is sufficient to perform a trace over the product of
Dirac matrices to obtain (again up to finite parts)
\begin{eqnarray}
(7c_1)+(7c_2)+(7c_3)
&=& 6g^4\mu^{6-D^\prime-D}\int {d^{D^\prime}k\over (2\pi)^{D^\prime}}
                     {d^Dq\over (2\pi)^D}
             {k^2q^2-(k\cdot q)^2
        \over k^2q^2(k+p)^2(q+p)^2(k+q+p)^2}\nonumber\\
&=& {3\over 8}g^4\mu^{3-D}\int {d^Dq\over (2\pi)^D}
             {1\over q^2|q+p|}\;,
\end{eqnarray}
where we have used
$q\cdot k = {1\over 2}{\bf [}(q+k+p)^2+p^2-(q+p)^2-(k+p)^2 {\bf ]}$.

Finally, upon using the formula
\begin{equation}
\mu^{3-D}\int {d^Dk\over (2\pi)^D}{1\over k(k+p)^2}
= {1\over 4\pi^2}{\bf [}{2\over 3-D}
    +ln({\mu^2\over p^2}){\bf ]}\;,
\end{equation}
and putting everything together, we obtain the following
divergent contribution to
the Fermion self-energy with mass insertion:
\begin{eqnarray}
\Gamma^{(2,0)}_{\bar{\psi}\psi}
= 1+ {5g^4\over 16\pi^2}{\bf [}{2\over 3-D}
    +ln({\mu^2\over p^2}){\bf ]}\;.
\label{g2}
\end{eqnarray}
Upon using the renormalization relation
\begin{equation}
(\Gamma^{(2,0)}_{{\bar{\psi}}\psi})_R
  = Z_\psi Z_{{\bar{\psi}}\psi}
\Gamma^{(2,0)}_{{\bar{\psi}}\psi} \;,
\end{equation}
and Eqs. (\ref{zpsi}), (\ref{g2}), we can obtain
the renormalization constant and the anomalous dimension
of the composite operator:
\begin{equation}
Z_{\bar{\psi}\psi} = 1- {17g^4\over 24\pi^2}{1\over 3-D}\;,
\end{equation}
\begin{equation}
\gamma_{\bar{\psi}\psi} \equiv
        {\partial Z_{\bar{\psi}\psi}\over \partial (1/\epsilon)}
       = -{17g^4\over 24\pi^2}\;.
\label{anom2}
\end{equation}
Finally, upon inserting Eqs. (\ref{anom1}) and
(\ref{anom2}) into (\ref{expid}),
which defines the
critical exponents $\eta$ and $\nu$, we find
up to second order in $\alpha$:
\begin{equation}
\eta = 1 - {1\over 3}\alpha^2 \; ,\;\;\;\;
\nu^{-1} = 1 + {17\over 6}\alpha^2 \; .
\end{equation}

3) 1/N Expansion

In this subsection we consider a large N expansion
which has the advantage of giving a non-vanishing leading
correction to the Fermion longitudinal conductivity,
in contrast to that found above.
To this end, we consider generalizing the Lagrangian
in Eqn.
(\ref{action1})
for the massless
Dirac field coupled to Chern-Simons gauge-field
to include N massless
Dirac fields.
Once again we ignore for now
the trivial contributions from the massive field.
The appropriate Lagrangian is
\begin{equation}
L= \sum_{i=1}^N \psi^{\dagger}_i
[\gamma_\mu (\partial_\mu - i{g\over \sqrt{N}} a_\mu) + M]\psi_i
+i{1\over 2} \epsilon_{\mu\nu\lambda}
a_\mu \partial_\nu a_\lambda.
\label{action2}
\end{equation}
where as usual the coupling constant has been scaled
by 1/N.
The $1/N$ expansion is a formal summation of
diagrams in powers of
$1/N$, rather than powers of $g^2$ as in the previous section.
Below we obtain the leading 1/N corrections to the conductivities
at the transition and the critical exponents.

At the critical point, $M=0$, the Feynman rules
in the Landau gauge now read
\begin{eqnarray}
S_0(p) &=& {1\over i\not\!p}\;,\\
G_0^{\mu\nu}(p) &=& -
   {\epsilon^{\mu\nu\lambda}p^\lambda\over p^2}\;,\label{gbare}\\
\Gamma_0^\mu (p) &=& i{g\over \sqrt{N}}\gamma^\mu\;.\label{Nf}
\end{eqnarray}
However, in the $1/N$ expansion, rather than using the `bare'
gauge field propogator,
Eq.\ (\ref{gbare}), it is more convenient to first sum up
the chain of Fermion bubble diagrams
shown in Figure. 6, since they each
contribute at the same order, namely
${\cal O}(1/N^0)$.
(Each fermion loop now carries an extra factor $N$
because of the summation over flavors, and the two interaction
vertices carry $g^2/N$.)  Upon using
Eqs.\ (\ref{pi}) and (\ref{pi1}) to help
sum up the bubbles in Fig.~$8$, we obtain an effective (or dressed)
gauge-field propagator,
denoted by a cross line in the Figures,
as follows
\begin{eqnarray}
& & G_{eff}^{\mu\nu}(p) = A{\delta^{\mu\nu}p^2-p^\mu p^\nu\over p^3}
     + B{\epsilon^{\mu\nu\lambda}p^\lambda \over p^2}\;,\label{geff}\\
& & A = {g^2\over 16}{1\over 1+({g^2\over 16})^2}\;,~~~~
B = -{1\over 1+({g^2\over 16})^2}\;.
\label{AB}
\end{eqnarray}

A feature of the $1/N$ expansion in the present model
is that non-trivial corrections to the critical exponents
arise already at leading order (in 1/N),
in contrast to the
weak coupling expansion, where it was necessary to go to
two-loop order to see corrections.
In addition, as verified by explicit
calculations below, the $1/N$ expansion maintains conformal
invariance at the transition point $M=0$, at least at the
leading order in $1/N$.

To leading order,
${\cal O}(1/N)$, the Fermion self-energy is given
by the diagram in
Fig.~$9$. We have verified that it does not shift
(or renormalize) the Fermion mass from $M=0$.
To extract the term that
is proportional to $\not\! p$, we perform
$-\frac{1}{2p^2}\not\!p{\bf Tr}\not\!p$ on the diagram in
Fig.~$9$. Keeping
only the divergent part, we thereby find
\begin{eqnarray}
\Sigma(p)&=& = i\not\!p{g^2 A \over 6\pi^2N}
    {\bf [}{2\over 3-D}+ln({\mu^2\over p^2}){\bf ]}\;.
\end{eqnarray}
which gives directly the anomalous dimension:
\begin{eqnarray}
\gamma_\psi = - {g^2 \; A \over 6\pi^2} {1\over N}\;.
\label{anom3}
\end{eqnarray}

Upon insertion of the operator
$\bar{\psi}\psi$ into the one-loop
Fermion self-energy, we have the three diagrams in Fig.~$10$,
which up to finite parts give a contribution:
\begin{eqnarray}
(10a) &=&
 2A{g^2\over N}\mu^{3-D}\int{d^Dk\over (2\pi)^D}
          {1\over (k+p)^2k}\nonumber\\
(10b)+(10c) &=&  -{g^4\over 2N}(A^2-B^2)\mu^{3-D}
  \int{d^Dk\over (2\pi)^D} {1\over (k+p)^2k}\;.
\end{eqnarray}
Performing the loop integrals then gives
\begin{eqnarray}
\Gamma^{(2,0)}_{\bar{\psi}\psi} &=& 1+{\bf [}2A-{(A^2-B^2)g^2\over 2}
    {\bf ]} {g^2\over 4\pi^2N}
    {\bf [}{2\over 3-D}+ln({\mu^2\over p^2}){\bf ]}\;,\\
Z_{\bar{\psi}\psi} &=& 1 - {\bf [}{8\over 3}A-
     {(A^2-B^2)g^2\over 2}{\bf ]}
     {g^2\over 2\pi^2N} {1\over 3-D}\;,
\end{eqnarray}
and the anomalous dimension
\begin{equation}
\gamma_{\bar{\psi}\psi} =
     - {\bf [}{8\over 3}A-{(A^2-B^2)g^2\over 2}{\bf ]}
     {g^2\over 2\pi^2N}\;.
\label{anom4}
\end{equation}

Finally, upon inserting Equations (\ref{anom3}) and (\ref{anom4})
into (\ref{expid}) we obtain
the leading 1/N
expressions for the critical exponents:
\begin{eqnarray}
\eta &=& 1 - {8\over 3}
      {\alpha^{2}\over 64+(\pi \alpha)^{2}} {1\over N}\; ,\\
\nu^{-1} &=& 1+\frac{128}{3}
   \frac{(128-(\pi\alpha)^2)
   \alpha^2}{(64+(\pi\alpha)^2)^2}\frac{1}{N} \; .
\end{eqnarray}

Next we calculate the
order $1/N$ corrections to the conductivities at the
critical point.
The relevant Feynman diagrams which contribute to the
Fermion polarization tensor
are listed in Figs.~$11$ and $12$. The diagram in Fig.~$12$
is zero by Furry's theorem, because it contains a closed Fermion loop
attached to an odd number of gauge-field lines.
(Essentially this is a consequence of charge-conjugation
invariance, since the Chern-Simons gauge boson has
odd charge parity.) Fig.~$11a$, $11b$, and $11c$ are linear in
$G_{eff}^{\mu\nu}$ and therefore linear in
$\epsilon^{\mu\nu\lambda}$ and in
$(\delta^{\mu\nu}-p^\mu p^\nu/p^2)$. Thus the odd contribution,
$\Pi_o(p)$,
can be read off directly from (\ref{piox}):
\begin{equation}
\Pi_{o(a+b+c)}(p)=
 {g^4 B\over 16\pi^{2}}(1+{\pi^{2}\over 4}){1\over N}\;.
\label{pioxx}
\end{equation}
Calcualting the even contribution,
$\Pi_e(p)$, however, is much more complicated.
In addition to the cumbersome
Dirac trace,
the evaluation of the Feynman integrals are highly
non-trivial. We leave the details to an appendix and
here only quote the final result,
\begin{equation}
\Pi_e(p)
         \approx -{3A\over 16}{g^4\over (2\pi)^2}{1\over N} \; .
\end{equation}
It should be emphasized that all divergent contributions to the
polarization tensor
$\Pi_{\mu\nu}(p)$ cancel, so that in the $1/N$ expansion
the $\beta$-function $\beta(g)$ vanishes, and conformal
invariance survives quantum fluctuations at the transition point.

Finally we can obtain the Fermion conductivities at the critical point
from the above polarization tensors.
Eq. (13) can be used to obtain the conductivity per flavor of massless
Dirac field.  To obtain the total Fermion conductivity per flavor
we must add to
$\sigma_{F,xy}$
the
contribution of -1/2
from the massive Dirac field.  The final result to leading order in 1/N
is:
\begin{eqnarray}
\sigma_{F,xy}
   &=& - {1 \over 2} + 16(1+{\pi^2\over 4})
{\alpha\over 64+(\pi\alpha)^{2}}{1\over N} \; ,\\
\sigma_{F,xx}
   &\approx& {\pi\over 8} (1 +
  {12\alpha^2 \over 64+(\pi\alpha)^2}{1\over N}) \;.
\end{eqnarray}

IV. DISCUSSION

As we have seen, the critical properties of the Mott anyon
transition vary continuously with the anyon statistics.
Thus the model describes a line of fixed points which are
characterized by the statistics parameter $\alpha$,
which varies from Fermions, $\alpha =0$, to Bosons,
$\alpha =1$.
In Section III we calculated the critical exponents and
universal conductivities at the anyon Mott transition
as an expansion around the Fermion point.
Specifically, in terms of the deviation from Fermi statistics,
$\alpha$, we obtained critical exponents up to second
order and conductivities to first order.
The critical properties at the Mott transition
can also be obtained
at $\alpha = 1$ directly in
terms of a Bosonic scalar (U(1))
field theory, or equivalently the 3D XY model \cite{FWGF}.
The exponents are of course
known quite accurately for the 3D XY model,
and recently an estimate for
the conductivity has been obtained
from Monte Carlo simulations
and a large N expansion \cite{MCXY}.
Also, within a large N calculation
Wen and Wu \cite{WW} have recently
performed an expansion for $\alpha$ near one by coupling
a Chern-Simons gauge-field to the U(1) Boson field.
It is instructive to compare these various results for the
critical properties of the Mott transition in order
to see the trends in exponents and conductivities
as one varies the particle statistics from
Fermi to Bose.  As we discuss below, the observed trend can
give one some insight into possible connections between
transitions in the integer quantum Hall effect \cite{QHEEXPT},
and the
magnetic field-tuned superconductor insulator transition in
thin films \cite{SIEXPT}, \cite{SITHEOR}.

Consider first the exponents $\eta$ and $\nu$.  In Section III
we found that to second order in $\alpha$,
\begin{equation}
\eta (\alpha) = 1 - {1\over 3}\alpha^2 +O(\alpha^3) \; ,\;\;\;\;
\nu^{-1} (\alpha) = 1 + {17\over 6}\alpha^2 + O(\alpha^3) \; .
\end{equation}
For the 3D XY model, which corresponds to $\alpha=1$,
the exponents are given by roughly,
\begin{equation}
\eta (\alpha=1) \sim 1/50 \; ,\;\;\;\;
\nu^{-1} (\alpha=1) \sim  3/2 \; .
\end{equation}
Notice that both $\eta$ and $\nu$ are smaller for Bosons than for
Fermions.  Moreover, the expansion from the Fermion end indicates
that the initial deviations for small $\alpha$ are to reduce
the exponents, suggesting that $\eta(\alpha)$ and $\nu(\alpha)$
might be monotonically decreasing functions of $\alpha$.
Unfortunately, as is clear from this expansion, the leading order
term does not give a reliable estimate at $\alpha=1$.
This should be contrasted with the $\epsilon=4-D$
expansion for the XY model \cite{MEF}
which gives reasonable exponent
values for the 3D model when the low order results are
extrapolated to $\epsilon=1$.

It is also instructive to compare the 3D XY exponents with the
exponents obtained from the
large N Fermion expansion
in Section 3:
\begin{eqnarray}
\eta &=& 1 - {8\over 3}
      {\alpha^{2}\over 64+(\pi \alpha)^{2}} {1\over N}\; ,\\
\nu^{-1} &=& 1+\frac{128}{3}
   \frac{(128-(\pi\alpha)^2)
   \alpha^2}{(64+(\pi\alpha)^2)^2}\frac{1}{N} \; .
\end{eqnarray}
If we put $\alpha=1$ these become
\begin{eqnarray}
\eta (\alpha=1)  = 1 - (0.036...)(1/N) + O(1/N^2)\; ,\\
\nu^{-1} (\alpha=1) = 1 + (0.924...)(1/N) + O(1/N^2)  \; ,
\end{eqnarray}
which, when extrapolated to N=1, should become equal to
the 3D XY model exponents.  Once again, although the
leading term in the expansion has the `correct' sign,
the extrapolations to N=1 using only the first term are clearly
not very reliable.

Next we consider the universal conductivities at the Mott transition.
For Fermions, $\alpha=0$, we obtained in Section III
that $\sigma_{xx} =\pi/8$ and $\sigma_{xy} = -1/2$.
For the Boson Mott transition (in zero field)
$\sigma_{xy}=0$ and the recent numerical estimates \cite{MCXY}
give $\sigma_{xx}(\alpha=1) =0.285 \pm 0.02$,
a value slightly smaller
than in the Fermion case.  It is instructive to see if
the expansion about the Fermion point for small $\alpha$
gives the correct trends.  For this purpose it is both more natural
and easier to compare resitivities,
since as (14) shows the longitudinal
anyon resistivity is simply equal to the Fermion longitudinal
resistivity.  Upon inverting the perturbative results in Equation
(32) and (33) for the Fermion conductivity, we obtain, using (14),
an expansion for the anyon longitudinal resistivity:
\begin{equation}
\rho_{xx} (\alpha) = (0.9715...) + (2.084...)\alpha + O(\alpha^2)   \;.
\end{equation}
Simarlarly, the transverse anyon resistivity is given by:
\begin{equation}
\rho_{xy} (\alpha) = (1.237...) - (0.492...)\alpha + O(\alpha^2)   \;.
\end{equation}
The universal resistivities at $\alpha=1$ follow
from the Boson Monte Carlo simulations and are \cite{MCXY}
\begin{equation}
\rho_{xx} (\alpha=1) = 3.5 \pm 0.2\;,~~~~~~~~~~  \rho_{xy} (\alpha=1)=0  \;.
\end{equation}
Notice that the sign of the
leading small $\alpha$ correction to both
$\rho_{xx}$ and $\rho_{xy}$ are such that the values tend towards the
boson values in (75).
This result suggests that,
just as with the critical exponents, the components of the
universal
resistivity tensor
vary monotonically upon moving along the fixed line
of critical points from Fermion to Boson statistics.

One might be tempted to compare directly
our large-N results to those obtained in
Ref. \cite{WW} for N-flavors of Bosons
coupled to the Chern-Simons
field.
However, the two large N-theories are probably not
continuously connected to one another (upon varying $\alpha$)
since the extension to $N$ fields brings into the
Lagrangian of the theory a new $SU(N)$ symmetry, which is
spontaneously broken upon crossing the Bosonic transition,
while unaffected across the Fermionic phase transition.

We now turn to the relevance of the results obtained in this
paper to two-dimensional
experimental systems which exhibit zero temperature
quantum phase transitions.  Unfortunately
we cannot make direct contact with
experiment, since our
simplified lattice model ignores both disorder and long-ranged
Coulomb interactions.  Nevertheless, the notion of a fixed line of
critical points parameterized by the statistics of the underlying
particles, along which the exponents vary continuosly, and
monotonically, is presumably rather more general.

To be specific, consider a more realistic model of anyons, with a
long-ranged Coulomb interaction moving in a quenched random potential
and external magnetic field.  In the Boson limit, $\alpha=1$,
as parameters are varied this model presumably undergoes
a transition from a superconducting phase to a localized Bose glass
insulator.  This transition is believed to be in the
appropriate universality class for real disordered
superconducting films, which are tuned with an external magnetic
field from the superconducting into insulating phases \cite{SITHEOR}.
This magnetic field tuned superconductor-insulator
transition has been recently studied experimentally in
considerable detail \cite{SIEXPT}.
In the Fermion limit, $\alpha=0$, with strong disorder the model
is presumed to exhibit transitions between integer quantum Hall
plateaus, for which there is also considerable experimental
data \cite{QHEEXPT}.
The critical properties of these two experimentally accessible
phase transitions are thus presumably
end points of a fixed line
of critical points which interpolates between them.
Moreover, transitions between fractional plateaus in
the quantum Hall effect are probably described correctly
by this same model with fractional statistics $\alpha$.
For example, the transition between the so-called Hall insulator
and the $\nu=1/3$ quantum Hall plateau can be described as
a condensation of fractional statistics particles with
$\alpha=2/3$.

Although at present we cannot calculate analytically
(or numerically) the critical properties along the fixed line
of these disordered anyon models, it is instructive to compare
the critical properties measured experimentally with the trends
obtained in our simplified clean lattice model.
For example, experiments \cite{QHEEXPT}
on the transition between integer
plateaus in the Hall effect find that the correlation
length exponent, $\nu$, times the dynamical exponent,
$z$, to be given by roughly 7/3.  Since one expects z=1
in the presence of 1/r Coulomb interactions \cite{FGG},
this gives
an estimate for $\nu \sim 7/3$.
At the magnetic field tuned
superconductor insulator transition, on the other hand, with z=1
assumed, the experiments give
a much smaller value, $\nu \sim 5/4$.
Thus upon
moving along the presumed line of fixed points
from the Fermion to Boson end,
the exponent $\nu$ apparently decreases.
It is noteworthy that this
same trend, a decreasing of $\nu$ moving from Fermion to Boson,
was what we found along the fixed line of Mott anyon transitions.
Perhaps it is generic that
Fermion transitions
are closer to their lower
critical dimension, with a larger $\nu$,
than their Boson counterparts.

It is also amusing to compare the experimental values
for the universal resistivites at the two transitions.
At the field tuned superconductor-insulator transition \cite{SIEXPT}
the longitudinal resistivity is found to cluster in the
range between 0.8 and 1.0, in units of $h/(2e)^2$,
where the Cooper pair has charge $2e$.
At the transition between integer plateaus in the quantum Hall
effect, experiments
find \cite{TSUI}
values of $\sigma_{xx}$ of roughly 0.2.
Combining this with $\sigma_{xy}=1/2$ gives
$\rho_{xx}$ of roughly 0.7, a value
slightly smaller than at the superconductor-insulator transition.
It is interesting that at the anyon Mott transition we
also find the longitudinal resistivity increases
moving from the Fermion to Boson end.  It is of course unclear
whether or not this trend is a generic property of
anyon phase transitions.

Finally, it is worth re-empasizing that the
existence of a fixed line
of critical points characterized by the anyon statistics
in our simple lattice anyon model, underscores the importance
of the fluctuating gauge field.  At a critical point, or more
generally in a gapless phase, the fluctuating Chern-Simons gauge field
cannot be simply thrown away, as in
the `flux-smearing' mean field approaches.
More specifically, we expect that the presence of
this fluctuating gauge field will most likely
make the transition between plateaus in the fractional quantum
Hall effect in a different universality class from that between integer
plateaus.

ACKNOWLEDGMENTS

W.C. thanks I. Affleck and G.W. Semenoff for discussions.
M.P.A.F. is extremely grateful to Nick Read and R. Shankar
for numerous clarifying conversations during the formative
part of this work, and to the Institute for Theoretical Physics
in Santa Barbara where part of this work was initiated.
Y.S.W. thanks X.G. Wen for helpful discussions, and R.A. Webb
and D.H. Lee for warm hospitality at IBM, and R. Jackiw and
X.G. Wen for warm hospitality at MIT.
This work was supported
in part by the Natural Sciences and Engineering Research Council
of Canada and U.S. National Science Foundation through grant
No. PHY-9008452 and by the National Science Foundation under grant
No. Phy89-04035.

APPENDIX:  Calculation of $\Pi_e(p)$ at order ${\cal O}(1/N)$

In this appendix we present the details for the two-loop
calculation of the even part of the gauge-boson self-energy
$\Pi_e(p)$ in the $1/N$ expansion.

Some useful formula for the $2\times 2$ $\gamma$-matrices are
\begin{eqnarray}
{\bf Tr}(\gamma^\mu\gamma^\nu\gamma^\lambda\gamma^\sigma) &=&
2 (\delta^{\mu\nu}\delta^{\lambda\sigma}
   +\delta^{\mu\sigma}\delta^{\nu\lambda}
   -\delta^{\mu\lambda}\delta^{\nu\sigma})\;,\\
{\bf Tr}(\gamma^\mu \gamma^\nu \gamma^\lambda)
&=& 2\epsilon^{\mu\nu\lambda}\;,\\
\gamma^\mu \gamma^\mu &=& -3\cdot {\bf 1}\;,\\
\gamma^\mu\gamma^\lambda\gamma^\mu &=& \gamma^\lambda\;,\\
\gamma^\mu\gamma^\sigma\gamma^\lambda\gamma^\mu &=&
2\delta^{\sigma\lambda} {\bf 1} -\gamma^\lambda\gamma^\sigma\;,\\
\not\!k\not\!k &=& -k^2{\bf 1}\;,\\
\gamma^\mu\gamma^\eta\gamma^\nu\epsilon^{\nu\mu\tau}p^\tau &=&
                           2p^\eta{\bf 1}\;.
\end{eqnarray}

The relevant diagrams are those in Fig.~$11$. Note that
$\Pi_e={1\over 2p}\delta^{\mu\nu}\Pi^{\mu\nu}$. We have
$\Pi_{e}^{(a+b)}$ and $\Pi_{e}^{(c)}$ are therefore given by
\begin{eqnarray}
& &  -{A\over p}{g^4\over N}\mu^{6-D^\prime-D}\int
{d^{D^\prime}k\over (2\pi)^{D^\prime}} {d^Dq\over (2\pi)^D} {\bf Tr}
{\gamma^\mu(\not\!k +\!\not\!p)\gamma^\mu\not\!k
\gamma^\eta(\not\!k +\not\!q)\gamma^\sigma\not\!k
(\delta^{\sigma\eta}q^2-q^\sigma q^\eta)
\over k^4(k+p)^2(k+q)^2q^3}\nonumber\\
&=&-4{A\over p}{g^4\over N}\mu^{6-D^\prime-D}\int
{d^{D^\prime}k\over (2\pi)^{D^\prime}} {d^Dq\over (2\pi)^D}
{q\cdot(q+k)(2k\cdot pk\cdot q +k^2k\cdot q - k^2p\cdot q)
\over k^4(k+p)^2(k+q)^2q^3}\nonumber\\
\label{9ab}
\end{eqnarray}
and
\begin{eqnarray}
 & & - {A\over p}{g^4\over 2N}\mu^{6-D^\prime-D}\int
{d^{D^\prime}k\over(2\pi)^{D^\prime}}{d^Dq\over(2\pi)^D}
{\bf Tr}{\gamma^\mu(\not\!k +\not\!p)\gamma^\eta(\not\!k +\not\!q +\not\!p)
\gamma^\mu(\not\!k +\not\!q)\gamma^\sigma\not\!k
(\delta^{\sigma\eta}q^2-q^\sigma q^\eta)
\over k^2(k+p)^2(k+q)^2(k+q+p)^2q^3}\;\nonumber\\
 &=& - 2{A\over p}{g^4\over N}\mu^{6-D^\prime-D}\int
{d^{D^\prime}k\over(2\pi)^{D^\prime}}{d^Dq\over(2\pi)^D}
{G(k,q,p)
\over k^2(k+p)^2(k+q)^2(k+q+p)^2q^3}\;,
\label{9c}
\end{eqnarray}
respectively, where
\begin{eqnarray}
G(k,q,p) &=& k^2{\bf [}q^4-q^2p\cdot q -2q^2k\cdot q-2q^2k\cdot p
            -(q\cdot p)^2+q^2p^2-k^2q^2{\bf ]}\nonumber\\
         & & + (k\cdot q){\bf [} -2q^2k\cdot q +2(q\cdot p)(k\cdot p)
            -2q^2k\cdot p -2q^2q\cdot p -p^2k\cdot q{\bf]}\nonumber\\
         & & + q^2(k\cdot p){\bf [} q^2-q\cdot p -2k\cdot p{\bf]}\;.
\end{eqnarray}
Performing the convergent $k$-integrations in Eqs.
(\ref{9ab}, \ref{9c}) at $D^\prime=3$, we have
\begin{equation}
\Pi_{e(a+b)}={A\over p}{g^4\over 8N}\mu^{3-D}\int {d^Dq\over (2\pi)^D}
{\bf [}{2q\cdot p + p^2\over q^3|q+p|} -{q\cdot p\over q^2|q+p|p}
       {\bf ]}\;.
\label{pieab}
\end{equation}
\begin{equation}
\Pi_{e(c)}=-{A\over p}{g^4\over 8N}\mu^{3-D}\int {d^Dq\over (2\pi)^D}
{\bf [}{2q\cdot p + p^2\over q^3|q+p|} +{4 p\over q^2|q+p|}
+{1\over q|q+p|} +{2q^4+5p^2q^2+2p^4\over(q\cdot p)q^2|q+p|p}
      {\bf ]}\;.
\label{piec}
\end{equation}

Due to the factor $q\cdot p$ in the denominator of the
last term in (\ref{piec}), it is hard to obtain an
analytic expression with the dimensional regularization. So we
will set $D=3$ and introduce a momentum cut-off $\Lambda$ in
both (\ref{pieab}) and (\ref{piec}). We will see the
logarithmically divergent terms indeed cancel. Also
we will discard linearly divergent terms, whose appearance
is an artifact of this regularization by momentum-cutoff.
(Such divergences do not appear in a regularization that
preserves Lorentz invariance.) This results in a
finite contribution to $\Pi_e$ at the order ${\cal O}(1/N)$:
\begin{equation}
\Pi_{e(a+b+c)}=-{A\over p}{g^4\over 8N}\int^\Lambda {d^3q\over (2\pi)^3}
{\bf [} {q\cdot p \over q^2|q+p|p} + {4 p\over q^2|q+p|}
+{1\over q|q+p|} +{2q^4+5p^2q^2+2p^4\over(q\cdot p)q^2|q+p|p}
      {\bf ]}\;.
\label{piex}
\end{equation}
Except for a factor $-{A\over 8N}{g^4\over (2\pi)^2}$,
the first three terms are, respectively,
\begin{equation}
-{2\over 9} - {1\over 3}ln{\Lambda^2\over p^2}\;,~~~~~~
8 + 4ln{\Lambda^2\over p^2}\;,~~~~~~ -2\;.
\end{equation}
The fourth term is a bit messy:
\begin{eqnarray}
& & {1\over 2}\int_0^1 dy {2+5y+2y^2\over y\sqrt{1+y}}
          ln{\sqrt{1+y}-\sqrt{y}\over \sqrt{1+y}+\sqrt{y}}
           + {1\over 2}\int_1^{{\Lambda^2\over p^2}} dy
             {2+5y+2y^2\over y\sqrt{1+y}}
          ln{\sqrt{1+y}-1\over \sqrt{1+y}+1}\nonumber\\
    &=& {1\over 2}\int_0^1 dy {2+5y+2y^2\over y\sqrt{1+y}}
          ln{\sqrt{1+y}-\sqrt{y}\over \sqrt{1+y}+\sqrt{y}}
      + C - {22\over 6}ln{\Lambda^2\over p^2}\;.
\label{term4}
\end{eqnarray}
where
\begin{equation}
  C= -{64\over 9}-{1\over 2}
        {\bf [}ln{\sqrt{2}+1\over\sqrt{2}-1}{\bf ]}^2
+{13\sqrt{2}\over 3}ln{\sqrt{2}+1\over \sqrt{2}-1}
   = 2.137820914\;.
\end{equation}
The integral in (\ref{term4}) is numerically evaluated to be
$-6.405956452\cdots\;$. The logarithmic divergences cancel out
and the final finite result is
\begin{equation}
\Pi_e(p) = -1.509642238{A\over 8N}{g^4\over (2\pi)^2}
         \approx -{3A\over 16N}{g^4\over (2\pi)^2}\;.
\end{equation}

\newpage

FIGURES

\unitlength=1.00mm
\linethickness{0.5pt}
\thicklines
\vspace{0.5cm}

\begin{picture}(160.00,80.00)

\thicklines

\put(70.,48.00){\circle*{2.00}}
\put(70,48.0){\line(-1,0){40.00}}
\put(70.2,48.0){\line(0,-1){40.00}}
\put(69.8,48.0){\line(0,-1){40.00}}
\put(70,48.0){\line(0,-1){40.00}}
\put(58,36.0){\line(1,1){12.00}}
\put(70,48.0){\vector(-1,-1){28.00}}
\put(42,20.0){\line(-1,-1){12.00}}
\put(30.,8.00){\circle*{2.00}}
\put(30,8.0){\line(1,0){40.00}}
\put(30,8.0){\line(0,1){40.00}}
\put(30.,48.00){\circle*{4.00}}
\put(30,48.0){\vector(1,-1){28.00}}
\put(58,20.0){\line(1,-1){12.00}}
\put(70.,8.00){\circle*{4.00}}
\put(27.0,5.00){\makebox(0,0)[r]{$1$}}
\put(27.0,51.00){\makebox(0,0)[r]{$2$}}
\put(73.0,51.00){\makebox(0,0)[l]{$3$}}
\put(73.0,5.00){\makebox(0,0)[l]{$4$}}
\put(27.0,28.00){\makebox(0,0)[r]{$t$}}
\put(50.0,51.00){\makebox(0,0)[l]{$t$}}
\put(73.0,28.00){\makebox(0,0)[l]{$-t$}}
\put(50.0,5.00){\makebox(0,0)[l]{$t$}}
\put(100,40.0){\vector(1,0){8.00}}
\put(100,40.0){\vector(0,1){8.00}}
\put(111.0,40.00){\makebox(0,0)[l]{$x$}}
\put(100.0,51.0){\makebox(0,0)[l]{$y$}}
\put(100.,30.00){\circle*{4.00}}
\put(100.,20.00){\circle*{2.00}}
\put(112.0,30.00){\makebox(0,0)[l]{$m$}}
\put(112.0,20.00){\makebox(0,0)[l]{$-m$}}
\put(98,10.0){\vector(1,0){6.00}}
\put(104,10.0){\line(1,0){3.00}}
\put(110.0,10.00){\makebox(0,0)[l]{$it^\prime/4$}}


\end{picture}

\begin{description}
\item[Fig. 1]
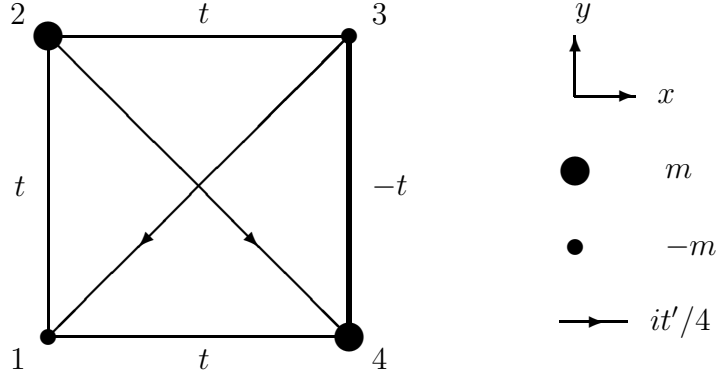
\ \ \ \ The four sites in the unit cell of the two-dimensional
square lattice tight binding model with one-half of a flux
quanta per plaquette and a staggered periodic potential.
Near neighbors have hopping strength $t$ except
the bold line which is $-t$.  The next-near neighbors have
a hopping strength $it^\prime /4$, and the staggered on-site
potential has strength $m$.
\end{description}

\begin{picture}(160.00,50.00)

\thicklines

\put(13.,18.00){\circle*{2.00}}
\put(20.0,18.00){\circle{40.00}}
\put(15.0,13.00){\line(1,1){10.00}}
\put(13.60,14.950){\line(1,1){9.00}}
\put(12.90,18.0){\line(1,1){7.00}}
\put(17.0,11.50){\line(1,1){9.00}}
\put(27.10,18.){\line(-1,-1){7.00}}
\put(27.,18.00){\circle*{2.00}}
\put(31.0,18.00){\makebox(0,0)[l]{$+$}}
\put(38.,18.00){\circle*{2.00}}
\put(45.0,18.00){\circle{40.00}}
\put(40.0,13.00){\line(1,1){10.00}}
\put(38.60,14.950){\line(1,1){9.00}}
\put(37.90,18.0){\line(1,1){7.00}}
\put(42.0,11.50){\line(1,1){9.00}}
\put(52.10,18.){\line(-1,-1){7.00}}
\multiput(52.00,18.0)(2.00,0.00){5}{\line(3,0){1.00}}
\put(68.0,18.00){\circle{40.00}}
\put(63.0,13.00){\line(1,1){10.00}}
\put(61.60,14.950){\line(1,1){9.00}}
\put(60.90,18.0){\line(1,1){7.00}}
\put(65.0,11.50){\line(1,1){9.00}}
\put(75.10,18.){\line(-1,-1){7.00}}
\put(75.,18.00){\circle*{2.00}}
\put(79.0,18.00){\makebox(0,0)[l]{$+$}}
\put(85.5,18.00){\circle*{2.00}}
\put(92.5,18.00){\circle{40.00}}
\put(87.50,13.00){\line(1,1){10.00}}
\put(86.10,14.950){\line(1,1){9.00}}
\put(85.40,18.0){\line(1,1){7.00}}
\put(89.50,11.50){\line(1,1){9.00}}
\put(99.50,18.0){\line(-1,-1){7.00}}
\multiput(99.50,18.0)(2.00,0.00){5}{\line(3,0){1.00}}
\put(115.5,18.00){\circle{40.00}}
\put(110.50,13.00){\line(1,1){10.00}}
\put(109.10,14.950){\line(1,1){9.00}}
\put(108.40,18.0){\line(1,1){7.00}}
\put(112.50,11.50){\line(1,1){9.00}}
\put(122.60,18.){\line(-1,-1){7.00}}
\multiput(122.50,18.0)(2.00,0.00){5}{\line(3,0){1.00}}
\put(138.5,18.00){\circle{40.00}}
\put(133.50,13.00){\line(1,1){10.00}}
\put(132.10,14.950){\line(1,1){9.00}}
\put(131.40,18.0){\line(1,1){7.00}}
\put(135.50,11.50){\line(1,1){9.00}}
\put(145.60,18.){\line(-1,-1){7.00}}
\put(145.5,18.00){\circle*{2.00}}
\put(149.50,18.00){\makebox(0,0)[l]{$+\ldots$}}
\end{picture}

\begin{description}
\item[Fig. 2]
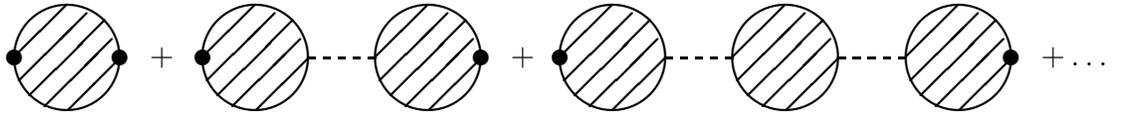
\ \ \ \ The full current-current correlation for the external
electromagnetic field. The black spots stand for the currents.
The striped disk represents the exact Chern-Simons self-energy
and the dashed line the Chern-Simons propagator.
\end{description}

\begin{picture}(110.00,62.00)
\multiput(37.00,18)(-2.00,0.00){3}{\line(3,0){1.00}}
\put(45.0,18.00){\circle{20.00}}
\multiput(52.00,18.0)(2.00,0.00){3}{\line(3,0){1.00}}
\put(45.0,4.00){\makebox(0,0)[cc]{$a$}}
\put(86.50,14.00){\line(1,0){26.00}}
\multiput(92.100,15.250)(1.00,2.00){4}{\line(3,0){1.00}}
\multiput(106.900,15.250)(-1.00,2.00){4}{\line(3,0){1.00}}
\multiput(97.50,22.90)(2.00,0.00){3}{\line(3,0){1.00}}
\put(100.00,4.00){\makebox(0,0)[cc]{$b$}}
\end{picture}
\begin{description}
\item[Fig. 3]
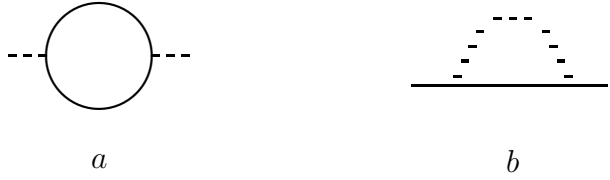
\ \ \ \ ($a$) Chern-Simons  and ($b$) fermion  self-energies
at ${\cal O}(e^2)$. The solid line stands for the fermion propagator.
\end{description}
%
%
%
\unitlength=1.0mm
\begin{picture}(110.00,80.00)
\thicklines
\put(20.0,20.00){\circle{30.00}}
\multiput(12.00,20)(-2.00,0.00){3}{\line(3,0){1.00}}
\multiput(27.00,20)(2.00,0.00){3}{\line(3,0){1.00}}
\multiput(14.50,16)(1.00,2.00){3}{\line(3,0){1.00}}
\multiput(24.50,16)(-1.00,2.00){3}{\line(3,0){1.00}}
\multiput(18.50,21)(2.00,0.00){2}{\line(3,0){1.00}}
\put(70.0,20.00){\circle{30.00}}
\multiput(62.00,20)(-2.00,0.00){3}{\line(3,0){1.00}}
\multiput(77.0,20)(2.00,0.00){3}{\line(3,0){1.00}}
\multiput(64.50,24.)(1.00,-2.00){3}{\line(3,0){1.00}}
\multiput(74.750,24.)(-1.00,-2.00){3}{\line(3,0){1.00}}
\multiput(68.650,19.1)(2.00,0.00){2}{\line(3,0){1.00}}
\put(120.0,20.00){\circle{30.00}}
\multiput(112.00,19.7)(-2.00,0.00){3}{\line(3,0){1.00}}
\multiput(127.00,20)(2.00,0.00){3}{\line(3,0){1.00}}
\multiput(120.00,25.50)(0.00,-2.00){7}{\line(0,3){1.00}}
\put(20.00,5.0){\makebox(0,0)[cc]{$a$}}
\put(70.00,5.0){\makebox(0,0)[cc]{$b$}}
\put(120.00,5.0){\makebox(0,0)[cc]{$c$}}

\end{picture}
\begin{description}
\item[Fig. 4]
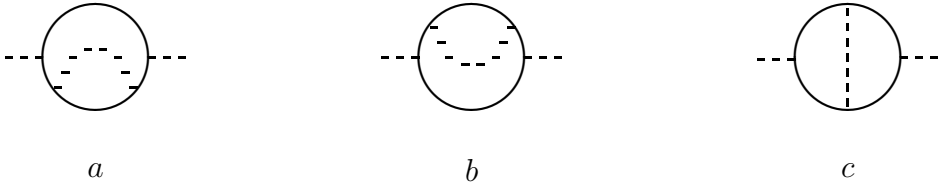
\ \ \ \ Chern-Simons self-energy at ${\cal O}(e^4)$.
\end{description}
\begin{picture}(110.00,80.00)
\thicklines
\put(5.00,12.50){\line(1,0){32.00}}
\multiput(10.00,14.00)(1.00,2.00){4}{\line(3,0){1.00}}
\multiput(31.00,14.00)(-1.00,2.00){4}{\line(3,0){1.00}}
\multiput(15.00,21.00)(-2.00,0.00){1}{\line(3,0){1.00}}
\multiput(26.00,21.00)(2.00,0.00){1}{\line(3,0){1.00}}
\put(21.0,21.00){\circle{10.00}}
\put(20.00,4.00){\makebox(0,0)[cc]{$a$}}
%
%
\put(54.00,18.0){\line(1,0){33.00}}
\multiput(59.00,19.50)(1.00,2.00){3}{\line(3,0){1.00}}
\multiput(81.50,19.50)(-1.00,2.00){3}{\line(3,0){1.00}}
\multiput(63.00,25.00)(2.00,0.00){8}{\line(3,0){1.00}}
\multiput(64.750,16.50)(1.00,-2.00){3}{\line(3,0){1.00}}
\multiput(75.20,16.50)(-1.00,-2.00){3}{\line(3,0){1.00}}
\multiput(69.00,11.00)(2.00,0.00){2}{\line(3,0){1.00}}
\put(70.00,4.00){\makebox(0,0)[cc]{$b$}}
%
%
\put(104.00,18.0){\line(1,0){33.00}}
\multiput(109.00,19.500)(1.00,2.00){3}{\line(3,0){1.00}}
\multiput(123.50,19.500)(-1.00,2.00){3}{\line(3,0){1.00}}
\multiput(113.00,25.00)(2.00,0.00){4}{\line(3,0){1.00}}
\multiput(116.750,16.50)(1.00,-2.00){3}{\line(3,0){1.00}}
\multiput(131.0,16.50)(-1.00,-2.00){3}{\line(3,0){1.00}}
\multiput(121.00,11.00)(2.00,0.00){4}{\line(3,0){1.00}}
\put(120.00,4.00){\makebox(0,0)[cc]{$c$}}

\end{picture}

\begin{description}
\item[Fig. 5]
\ \ \ \ Fermion self-energy at ${\cal O}(e^4)$.

\end{description}
\begin{picture}(110.00,80.00)
\thicklines
\multiput(12.00,18)(-2.00,0.00){3}{\line(3,0){1.00}}
\put(20.0,18.00){\circle{20.00}}
\multiput(27.00,18.0)(2.00,0.00){3}{\line(3,0){1.00}}
\put(20.0,4.00){\makebox(0,0)[cc]{$a$}}
\put(20.00,25.0){\makebox(0,0)[cc]{$\times$}}
\multiput(62.00,18)(-2.00,0.00){3}{\line(3,0){1.00}}
\put(70.0,18.00){\circle{20.00}}
\multiput(77.00,18.0)(2.00,0.00){3}{\line(3,0){1.00}}
\put(70.0,4.00){\makebox(0,0)[cc]{$b$}}
\put(70.00,11.0){\makebox(0,0)[cc]{$\times$}}
\put(106.00,14.00){\line(1,0){27.00}}
\multiput(111.00,15.50)(1.00,2.00){4}{\line(3,0){1.00}}
\multiput(127.00,15.50)(-1.00,2.00){4}{\line(3,0){1.00}}
\multiput(116.00,23.00)(2.00,0.00){4}{\line(3,0){1.00}}
\put(119.500,4.00){\makebox(0,0)[cc]{$c$}}
\put(119.500,14.0){\makebox(0,0)[cc]{$\times$}}

\end{picture}

\begin{description}
\item[Fig. 6]
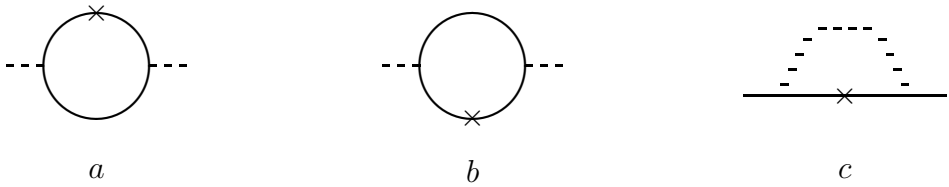
\ \ \ \ Vertices with composite operator
insertions at ${\cal O}(e^2)$.
The cross stands for the composite operator
$\bar\psi\psi$.

\end{description}
\begin{picture}(110.00,100.00)
\thicklines
\put(5.00,92.50){\line(1,0){32.00}}
\multiput(10.00,94.00)(1.00,2.00){4}{\line(3,0){1.00}}
\multiput(31.00,94.00)(-1.00,2.00){4}{\line(3,0){1.00}}
\multiput(15.00,101.00)(-2.00,0.00){1}{\line(3,0){1.00}}
\multiput(26.00,101.00)(2.00,0.00){1}{\line(3,0){1.00}}
\put(21.0,101.00){\circle{10.00}}
\put(21.00,84.00){\makebox(0,0)[cc]{$a_1$}}
\put(21.00,92.50){\makebox(0,0)[cc]{$\times$}}
\put(54.00,98.0){\line(1,0){33.00}}
\multiput(59.00,99.50)(1.00,2.00){3}{\line(3,0){1.00}}
\multiput(81.50,99.50)(-1.00,2.00){3}{\line(3,0){1.00}}
\multiput(63.00,105.00)(2.00,0.00){8}{\line(3,0){1.00}}
\multiput(64.750,96.50)(1.00,-2.00){3}{\line(3,0){1.00}}
\multiput(75.20,96.50)(-1.00,-2.00){3}{\line(3,0){1.00}}
\multiput(69.00,91.00)(2.00,0.00){2}{\line(3,0){1.00}}
\put(70.50,84.00){\makebox(0,0)[cc]{$b_1$}}
\put(70.5,98.0){\makebox(0,0)[cc]{$\times$}}
\put(104.00,98.0){\line(1,0){33.00}}
\multiput(109.00,99.500)(1.00,2.00){3}{\line(3,0){1.00}}
\multiput(123.50,99.500)(-1.00,2.00){3}{\line(3,0){1.00}}
\multiput(113.00,105.00)(2.00,0.00){4}{\line(3,0){1.00}}
\multiput(116.750,96.50)(1.00,-2.00){3}{\line(3,0){1.00}}
\multiput(131.0,96.50)(-1.00,-2.00){3}{\line(3,0){1.00}}
\multiput(121.00,91.00)(2.00,0.00){4}{\line(3,0){1.00}}
\put(120.50,84.00){\makebox(0,0)[cc]{$c_1$}}
\put(120.5,98.0){\makebox(0,0)[cc]{$\times$}}
\put(5.00,52.50){\line(1,0){32.00}}
\multiput(10.00,54.00)(1.00,2.00){4}{\line(3,0){1.00}}
\multiput(31.00,54.00)(-1.00,2.00){4}{\line(3,0){1.00}}
\multiput(15.00,61.00)(-2.00,0.00){1}{\line(3,0){1.00}}
\multiput(26.00,61.00)(2.00,0.00){1}{\line(3,0){1.00}}
\put(21.0,61.00){\circle{10.00}}
\put(21.00,44.00){\makebox(0,0)[cc]{$a_2$}}
\put(21.00,66.0){\makebox(0,0)[cc]{$\times$}}
\put(54.00,58.0){\line(1,0){33.00}}
\multiput(59.00,59.50)(1.00,2.00){3}{\line(3,0){1.00}}
\multiput(81.50,59.50)(-1.00,2.00){3}{\line(3,0){1.00}}
\multiput(63.00,65.00)(2.00,0.00){8}{\line(3,0){1.00}}
\multiput(64.750,56.50)(1.00,-2.00){3}{\line(3,0){1.00}}
\multiput(75.20,56.50)(-1.00,-2.00){3}{\line(3,0){1.00}}
\multiput(69.00,51.00)(2.00,0.00){2}{\line(3,0){1.00}}
\put(70.00,44.00){\makebox(0,0)[cc]{$b_2$}}
\put(62.10,58.0){\makebox(0,0)[cc]{$\times$}}
\put(104.00,58.0){\line(1,0){33.00}}
\multiput(109.00,59.500)(1.00,2.00){3}{\line(3,0){1.00}}
\multiput(123.50,59.500)(-1.00,2.00){3}{\line(3,0){1.00}}
\multiput(113.00,65.00)(2.00,0.00){4}{\line(3,0){1.00}}
\multiput(116.750,56.50)(1.00,-2.00){3}{\line(3,0){1.00}}
\multiput(131.0,56.50)(-1.00,-2.00){3}{\line(3,0){1.00}}
\multiput(121.00,51.00)(2.00,0.00){4}{\line(3,0){1.00}}
\put(120.00,44.00){\makebox(0,0)[cc]{$c_2$}}
\put(113.500,58.0){\makebox(0,0)[cc]{$\times$}}
\put(5.00,12.50){\line(1,0){32.00}}
\multiput(10.00,14.00)(1.00,2.00){4}{\line(3,0){1.00}}
\multiput(31.00,14.00)(-1.00,2.00){4}{\line(3,0){1.00}}
\multiput(15.00,21.00)(-2.00,0.00){1}{\line(3,0){1.00}}
\multiput(26.00,21.00)(2.00,0.00){1}{\line(3,0){1.00}}
\put(21.0,21.00){\circle{10.00}}
\put(20.00,4.00){\makebox(0,0)[cc]{$a_3$}}
\put(21.00,16.150){\makebox(0,0)[cc]{$\times$}}
\put(54.00,18.0){\line(1,0){33.00}}
\multiput(59.00,19.50)(1.00,2.00){3}{\line(3,0){1.00}}
\multiput(81.50,19.50)(-1.00,2.00){3}{\line(3,0){1.00}}
\multiput(63.00,25.00)(2.00,0.00){8}{\line(3,0){1.00}}
\multiput(64.750,16.50)(1.00,-2.00){3}{\line(3,0){1.00}}
\multiput(75.20,16.50)(-1.00,-2.00){3}{\line(3,0){1.00}}
\multiput(69.00,11.00)(2.00,0.00){2}{\line(3,0){1.00}}
\put(70.00,4.00){\makebox(0,0)[cc]{$b_3$}}
\put(79.0,18.0){\makebox(0,0)[cc]{$\times$}}
\put(104.00,18.0){\line(1,0){33.00}}
\multiput(109.00,19.500)(1.00,2.00){3}{\line(3,0){1.00}}
\multiput(123.50,19.500)(-1.00,2.00){3}{\line(3,0){1.00}}
\multiput(113.00,25.00)(2.00,0.00){4}{\line(3,0){1.00}}
\multiput(116.750,16.50)(1.00,-2.00){3}{\line(3,0){1.00}}
\multiput(131.0,16.50)(-1.00,-2.00){3}{\line(3,0){1.00}}
\multiput(121.00,11.00)(2.00,0.00){4}{\line(3,0){1.00}}
\put(120.00,4.00){\makebox(0,0)[cc]{$c_3$}}
\put(128.00,18.0){\makebox(0,0)[cc]{$\times$}}

\end{picture}

\begin{description}
\item[Fig. 7]
\ \ \ \
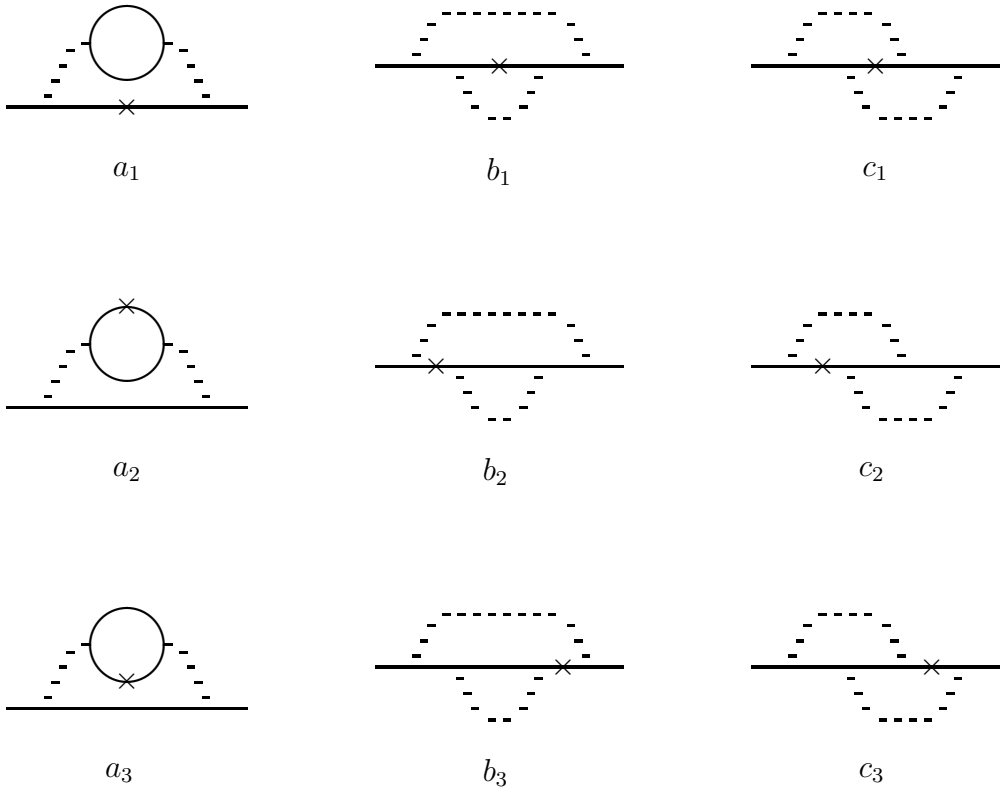
Fermion vertex with the insertion of the composite operator
$\bar\psi\psi$ at ${\cal O}(e^4)$.

\end{description}
\begin{picture}(170.00,62.00)

\thicklines
\multiput(0.00,18.)(2.00,0.00){10}{\line(3,0){1.00}}
\multiput(0.500,17.5)(2.00,0.00){10}{\line(0,3){1.00}}
\put(25.00,18.00){\makebox(0,0)[cc]{ = }}
\multiput(30.00,18)(2.00,0.00){9}{\line(3,0){1.00}}
\put(51.00,18.00){\makebox(0,0)[cc]{ + }}
\multiput(54.00,18)(2.00,0.00){5}{\line(3,0){1.00}}
\put(70.0,18.00){\circle{20.00}}
\multiput(77.00,18.0)(2.00,0.00){5}{\line(3,0){1.00}}
\put(89.00,18.00){\makebox(0,0)[cc]{ + }}
\multiput(92.00,18.0)(2.00,0.00){5}{\line(3,0){1.00}}
\put(108.0,18.00){\circle{20.00}}
\multiput(115.00,18.0)(2.00,0.00){4}{\line(3,0){1.00}}
\put(129.0,18.00){\circle{20.00}}
\multiput(136.00,18.00)(2.00,0.00){5}{\line(3,0){1.00}}
\put(151.00,18.00){\makebox(0,0)[cc]{ + \ $\ldots$}}

\end{picture}

\begin{description}
\item[Fig. 8]
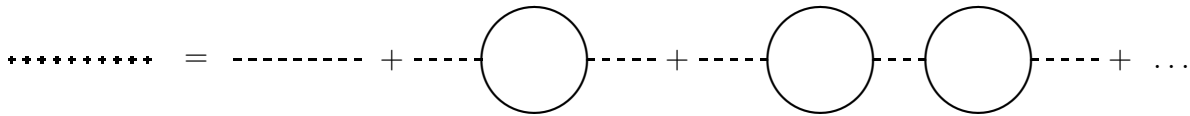
\ \ \ \ Summation of infinite series of one-loop Fermion
bubble chains gives the dressed Chern-Simons propagator, which
is of order ${\cal O}(N^0)$.
The cross line is the dressed Chern-Simons propagator, the
solid lines are the Fermion propagators, the dashed
lines are the bare Chern-Simons propagators.
\end{description}
\unitlength=1.00mm
\linethickness{0.5pt}
\vspace{0.5cm}
\begin{picture}(110.00,80.00)
\thicklines
\put(56.0,12.50){\line(1,0){27.00}}
\multiput(61.00,14.00)(1.00,2.00){4}{\line(3,0){1.00}}
\multiput(77.00,14.00)(-1.00,2.00){4}{\line(3,0){1.00}}
\multiput(66.00,21.00)(2.00,0.00){4}{\line(3,0){1.00}}

\multiput(61.500,13.500)(1.00,2.00){4}{\line(0,3){1.00}}
\multiput(77.500,13.500)(-1.00,2.00){4}{\line(0,3){1.00}}
\multiput(66.5,20.500)(2.00,0.00){4}{\line(0,3){1.00}}
%
\end{picture}
\begin{description}
\item[Fig. 9]
\ \ \ \ Fermion self-energy at ${\cal O}(1/N)$.
\end{description}
\begin{picture}(110.00,80.00)
\thicklines
\put(6.50,12.00){\line(1,0){27.00}}
\multiput(11.50,13.50)(1.00,2.00){4}{\line(3,0){1.00}}
\multiput(27.50,13.50)(-1.00,2.00){4}{\line(3,0){1.00}}
\multiput(16.50,21.00)(2.00,0.00){4}{\line(3,0){1.00}}
\multiput(12.00,13.00)(1.00,2.00){4}{\line(0,3){1.00}}
\multiput(28.00,13.00)(-1.00,2.00){4}{\line(0,3){1.00}}
\multiput(17.00,20.500)(2.00,0.00){4}{\line(0,3){1.00}}
\put(20.00,4.00){\makebox(0,0)[cc]{$a$}}
\put(20.0,12.0){\makebox(0,0)[cc]{$\times$}}
\put(54.00,12.00){\line(1,0){34.00}}
\multiput(59.50,13.50)(1.00,2.00){4}{\line(3,0){1.00}}
\multiput(81.50,13.50)(-1.00,2.00){4}{\line(3,0){1.00}}
\multiput(64.50,21.0)(-2.00,0.00){1}{\line(3,0){1.00}}
\multiput(76.50,21.0)(2.00,0.00){1}{\line(3,0){1.00}}
\multiput(60.00,13.00)(1.00,2.00){4}{\line(0,3){1.00}}
\multiput(82.00,13.00)(-1.00,2.00){4}{\line(0,3){1.00}}
\multiput(65.00,20.500)(-2.00,0.00){1}{\line(0,3){1.00}}
\multiput(77.00,20.500)(2.00,0.00){1}{\line(0,3){1.00}}
\put(71.0,21.0){\circle{10.00}}
\put(71.00,4.00){\makebox(0,0)[cc]{$b$}}
\put(71.00,26.0){\makebox(0,0)[cc]{$\times$}}
\put(104.00,12.00){\line(1,0){34.00}}
\multiput(109.50,13.50)(1.00,2.00){4}{\line(3,0){1.00}}
\multiput(131.50,13.50)(-1.00,2.00){4}{\line(3,0){1.00}}
\multiput(114.50,21.0)(-2.00,0.00){1}{\line(3,0){1.00}}
\multiput(126.50,21.0)(2.00,0.00){1}{\line(3,0){1.00}}
\multiput(110.00,13.00)(1.00,2.00){4}{\line(0,3){1.00}}
\multiput(132.00,13.00)(-1.00,2.00){4}{\line(0,3){1.00}}
\multiput(115.00,20.500)(-2.00,0.00){1}{\line(0,3){1.00}}
\multiput(127.00,20.500)(2.00,0.00){1}{\line(0,3){1.00}}
\put(121.0,21.0){\circle{10.00}}
\put(121.00,4.00){\makebox(0,0)[cc]{$c$}}
\put(121.00,16.0){\makebox(0,0)[cc]{$\times$}}

\end{picture}

\begin{description}
\item[Fig. 10]
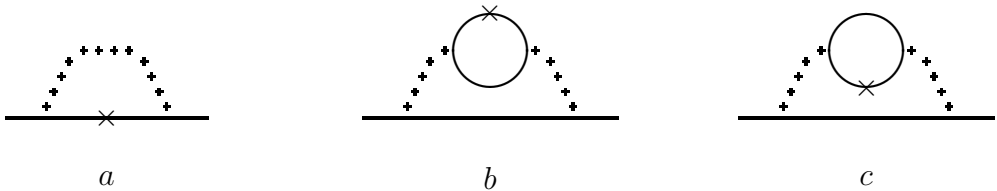
\ \ \ \ Fermion vertex with an insertion of the
 composite operator $\bar\psi_i\psi_i$ at ${\cal O}(1/N)$.
The cross stands for the composite operator  $\bar\psi_i\psi_i$.

\end{description}
\unitlength=1.0mm
\begin{picture}(110.00,80.00)
\thicklines
\put(20.0,20.00){\circle{30.00}}
\multiput(12.00,20)(-2.00,0.00){3}{\line(3,0){1.00}}
\multiput(27.00,20)(2.00,0.00){3}{\line(3,0){1.00}}
\multiput(14.50,16)(1.00,2.00){3}{\line(3,0){1.00}}
\multiput(24.750,16)(-1.00,2.00){3}{\line(3,0){1.00}}
\multiput(18.50,21)(2.00,0.00){2}{\line(3,0){1.00}}
\multiput(12.50,19.5)(-2.00,0.00){3}{\line(0,3){1.00}}
\multiput(27.50,19.5)(2.00,0.00){3}{\line(0,3){1.00}}
\multiput(15.0,15.5)(1.00,2.00){3}{\line(0,3){1.00}}
\multiput(25.250,15.5)(-1.00,2.00){3}{\line(0,3){1.00}}
\multiput(19.0,20.5)(2.00,0.00){2}{\line(0,3){1.00}}
\put(70.0,20.00){\circle{30.00}}
\multiput(62.00,20)(-2.00,0.00){3}{\line(3,0){1.00}}
\multiput(77.0,20)(2.00,0.00){3}{\line(3,0){1.00}}
\multiput(64.50,24.)(1.00,-2.00){3}{\line(3,0){1.00}}
\multiput(74.750,24.)(-1.00,-2.00){3}{\line(3,0){1.00}}
\multiput(68.650,19.1)(2.00,0.00){2}{\line(3,0){1.00}}
\multiput(62.50,19.5)(-2.00,0.00){3}{\line(0,3){1.00}}
\multiput(77.5,19.5)(2.00,0.00){3}{\line(0,3){1.00}}
\multiput(65.0,23.5)(1.00,-2.00){3}{\line(0,3){1.00}}
\multiput(75.250,23.5)(-1.00,-2.00){3}{\line(0,3){1.00}}
\multiput(69.150,18.6)(2.00,0.00){2}{\line(0,3){1.00}}
\put(120.0,20.00){\circle{30.00}}
\multiput(112.00,20)(-2.00,0.00){3}{\line(3,0){1.00}}
\multiput(127.00,20)(2.00,0.00){3}{\line(3,0){1.00}}
\multiput(119.50,26.0)(0.00,-2.00){7}{\line(3,0){1.00}}
\multiput(112.50,19.5)(-2.00,0.00){3}{\line(0,3){1.00}}
\multiput(127.50,19.5)(2.00,0.00){3}{\line(0,3){1.00}}
\multiput(120.00,25.50)(0.00,-2.00){7}{\line(0,3){1.00}}
\put(20.00,5.0){\makebox(0,0)[cc]{$a$}}
\put(70.00,5.0){\makebox(0,0)[cc]{$b$}}
\put(120.00,5.0){\makebox(0,0)[cc]{$c$}}

\end{picture}
\begin{description}
\item[Fig. 11]

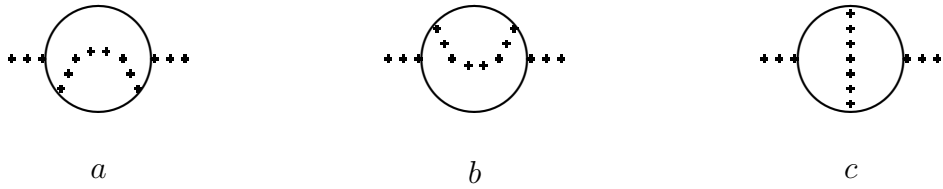
\ \ \ \ Chern-Simon self-energy at ${\cal O}(1/N)$.
\end{description}
%
%
%
\unitlength=0.90mm

\begin{picture}(110.00,75.00)

\thicklines

\put(65.0,20.00){\circle{30.00}}
\put(87.0,20.00){\circle{30.00}}
\multiput(56.00,20.)(-2.00,0.00){3}{\line(3,0){1.00}}
\multiput(72.50,23.5)(2.00,0.00){4}{\line(3,0){1.00}}
\multiput(72.50,16.5)(2.00,0.00){4}{\line(3,0){1.00}}
\multiput(95.00,20.)(2.00,0.00){3}{\line(3,0){1.00}}
\multiput(56.50,19.5)(-2.00,0.00){3}{\line(0,3){1.00}}
\multiput(73.0,23.)(2.00,0.00){4}{\line(0,3){1.00}}
\multiput(73.0,16.)(2.00,0.00){4}{\line(0,3){1.00}}
\multiput(95.50,19.5)(2.00,0.00){3}{\line(0,3){1.00}}

\end{picture}
\begin{description}
\item[Fig. 12]
\ \ \ \ Null diagram at order $O(1/N)$.
\end{description}


\begin{references}
%
\bibitem{STAT}
F. Wilczek, Phys.\ Rev.\ Lett.\  {\bf 48}, 1144 (1982);
Phys.\ Rev.\ Lett.\  {\bf 49} 957 (1982);
Y.S. Wu, Phys.\ Rev.\ Lett.\  {\bf 53}, 111 (1984).
%
\bibitem{FLH}
A. Fetter, C. Hanna and R. Laughlin,Phys. Rev. {\bf B39}
(1989) 9679.
%
\bibitem{GirMac}
S.M. Girvin and A.H. Macdonald, Phys.\ Rev.\ Lett.\ {\bf 58}
1252 (1987).
%
\bibitem{ZhHK}
S.C. Zhang, T.H. Hansen and S. Kivelson, Phys.\ Rev.\ Lett.\
{\bf 62}, 82 (1989);
D.H. Lee and S.C. Zhang, Phys.\ Rev.\ Lett.\  {\bf 66}, 1220 (1991).
%
\bibitem{Read}
N. Read, Phys.\ Rev.\ Lett.\  {\bf 62}, 86 (1989).
%
\bibitem{Jain}
J.K. Jain, Phys.\ Rev.\ Lett.\ {\bf 63} (1989) 199;
Phys. Rev.~ {\bf B40} (1990) 8079.
%
\bibitem{WB}
B. Blok and X.G. Wen, Phys. Rev.~ {\bf B42} (1990) 8133;
X.G. Wen and A. Zee, Nucl. Phys. {\bf B351} (1990) 135;
X.G. Wen, Mod. Phys. Lett. {\bf B5} (1991) 39.
%
\bibitem{HLR}
B.I. Halperin, P.A. Lee and N. Read, preprint, 1992.
%
\bibitem{LKZh}
D.H. Lee, S. Kivelson and S.C. Zhang, Phys.\ Rev.\ Lett.\
{\bf 68}, 2389 (1992); preprint, 1992.
%
\bibitem{NumDiag}
G.S. Canright, S.M. Girvin and A. Brass, Phys. Rev. Lett.
{\bf63}, 2295 (1989).
%
\bibitem{FWGF}
M.P.A. Fisher, P.B. Weichman, G. Grinstein and D.S. Fisher,
Phys. Rev B{\bf40}, 546 (1989).
%
\bibitem{MCXY}
M. Cha, M.P.A. Fisher, S.M. Girvin, M. Wallin and A.P. Young,
Phys. Rev B{\bf44}, 6883 (1991).
%
\bibitem{WW}
X.G. Wen and Y.-S. Wu, MIT preprint (July, 1992).
%
\bibitem{fluxphase}
A tight binding model with half of a flux quantum per plaquette
arises in `flux-phase' mean field treatments of frustrated
two-dimensional antiferromagnets.  See, for example,
X.G. Wen et. al. Phys Rev. B{\bf39}, 11413 (1989).
%
\bibitem{TKNN}
D.J. Thouless, M. Kohmoto, N.P. Nightingale and M. den Nijs,
Phys. Rev. Lett. {\bf49}, 405 (1982).
%
\bibitem{SHANKAR}
We are grateful to R. Shankar for the following particularly
succinct derivation of the final Dirac representation
in Equation (7).
%
\bibitem{CSL}
W. Siegel. Nucl. Phys. {\bf B156} (1979) 135; J.J. Schonfeld,
Nucl.Phys. {\bf B185} (1981) 157; R. Jackiw and S. Templeton,
Phys. Rev.~ {\bf D32} (1981) 2291.
%
\bibitem{FGG}
M.P.A. Fisher, G. Grinstein and S.M. Girvin, Phys. Rev. Lett.
{\bf64}, 587 (1990);  X.G. Wen and A. Zee, Int. J. Mod. Phys.B
{\bf4}, 437 (1990).
%
\bibitem{Amit}
D. Amit, {\it Field Theory, Renormalization Group and Critical
Phenomena},(Second Edition), World Scientific, 1988.
%
\bibitem{Parisi}
G. Parisi, {\it Statistical Field Theory}, Addison-Wesley, 1988.
%
\bibitem{Ramond}
P. Ramond, {\it A Priemer of Field Theory}, Addison-Wesley, 1988.
%
\bibitem{DJT}
S. Deser, R. Jackiw and S. Templeton, Phys. Rev. Lett.
{\bf 48} (1982) 975; Ann. Phys. (N.Y.) {\bf 140} (1982) 372.
%
\bibitem{CH}
S. Coleman and B. Hill, Phys. Lett. {\bf 159B} (1985) 184.
%
\bibitem{Redlich}
N. Redlich, Phys. Rev. {\bf D29} (1984) 2366.
%
\bibitem{NSW}
A. Niemi, G.W. Semenoff and Y.-S. Wu, Nucl. Phys.
{\bf B276} (1986) 173.
%
\bibitem{SSW}
G.W. Semenoff, P. Sodano and Y.-S. Wu, Phys.
Rev. Lett. {\bf 62} (1989) 715.
%
\bibitem{ChSW1}
W. Chen, G.W. Semenoff and Y.-S. Wu, Mod.\ Phys.\ Lett.\
{\bf A5 } (1990) 1833; in {\it Physics, Geometry and Topology},
Proc. of Banff Summer School on Particles and Fields, August
1989; (Plenum Pub. Cor.), p.553, 1990.
%
\bibitem{ChSW2}
W. Chen, G.W. Semenoff and Y.-S. Wu, Phys. Rev. {\bf D44} (1991)
R1625; and Phys. Rev. D, in press.
%
\bibitem{Ch1}
W. Chen, Phys. Lett. {\bf 251B} (1990) 415.
%
\bibitem{Ch2}
W. Chen (unpublished) (1990).
\bibitem{ST}
V.P. Spiridonov and F.V. Tkachov, Phys. Lett. {\bf B260} (1991) 109.
%
\bibitem{Other}
There is a huge literature for perturbative calculation of
the induced Chern-Simons term in various $D=2+1$ quantum field
theories other than Abelian Chern-Simons coupled to
non-self-interacting matter. An incomplete list is: Y. Kao and
M. Suzuki, Phys. rev. {\bf D31} (1985) 2137; M. Bernstien and
T.J. Lee, Phys. Rev.~ {\bf D32} (1985) 1020; R. Pisarski
and S. Rao, Phys. Rev.~ {\bf D32} (1985) 2081; T. Lee, Phys.
Lett. {171B} (1986) 247; V.P. Spidonov, JETP Lett. {\bf 52} (1990)
1112; L.V. Avdeev, G.V. Grigoriev and D.I. Kazakov, CERN preprint
TH-6091/91 (1991); G. Ferretti and S.G. Rajeev, Mod. Phys. Lett.
{bf A7} (1992) 2087; S.H. Park, Phys. Rev.~ {\bf D45} (1992) R3332; D.K. Hong,
T. Lee and S.H. Park, Korea preprint SNUTP 92/91 (1992); J. Chay,
D.K. Hong, T. Lee and S.H. Park, Korea preprint SNUTP 92/92 (1992).

%
\bibitem{NoteA}
In fact the beta-function $\beta (g)$ vanishes by explicit
calculation at two loops even in non-Abelian theory \cite{ChSW2};
it is proved to vanish up to all orders in perturbation theory
in massive Dirac theory, and is conjectured to do so in massless
theory. An argument for the latter case can be given as follows:
The Chern-Simons Lagrangian is gauge invariant only up to total
divergence terms. But on the other hand, only strictly gauge
invariant local counterterms can arise in a gauge-invariant
regularization and renormalization scheme. So no infinite counterterm
for the Chern-Simons Lagrangian can appear.
%
\bibitem{NoteB}
The divergence of the pole $2/\epsilon\equiv 2/(3-D)$ in the
physical dimension $D=3$ corresponds to $ln (\Lambda^2/\mu^2)$
in the regularization with momentum cutoff $\Lambda$.
%
\bibitem{QHEEXPT}
H.P. Wei, D.C. Tsui, M.A. Paalanen and A.M.M. Pruisken, Phys. Rev.
Lett. {\bf 61}, 1294 (1988).
%
\bibitem{SIEXPT}
A.F. Hebard and M.A. Paalanen, Phys. Rev. Lett. {\bf65}, 927
(1990).
%
\bibitem{SITHEOR}
M.P.A. Fisher, Phys. Rev. Lett. {\bf65}, 923 (1990).
%
\bibitem{MEF}
M.E. Fisher, Rev. Mod. Phys. {\bf46}, 597 (1974).
%
\bibitem{TSUI}
D.C. Tsui (private communication).
%
\end{references}
\end{document}